\documentclass[12pt]{article}
\usepackage{amsmath,amssymb}
\usepackage{amsthm}
\usepackage{mathtools}
\usepackage{lmodern}
\usepackage{iftex}
\usepackage[T1]{fontenc}
\usepackage{textcomp} 
\usepackage{graphicx} 
\usepackage[dvipsnames]{xcolor}
\usepackage[left=2.5cm,right=2.5cm,top=2.5cm,bottom=2.5cm]{geometry}
\usepackage{longtable,booktabs,array}
\usepackage{tikz}
\usetikzlibrary{matrix,shapes.geometric,calc}
\usepackage{amsmath}
\usepackage{soul}
\usepackage{yhmath}
\usepackage{csquotes}
\usepackage{rotating}
\usepackage{pdflscape}
\usepackage{eurosym}
\usepackage{booktabs}
\usepackage{threeparttable}
\usepackage{threeparttablex}
\usepackage{calc} 
\usepackage{hyperref}
\hypersetup{
  colorlinks=true,
  linkcolor=blue,
  citecolor=blue,
  filecolor=black,  
  urlcolor=blue,
  pdftitle={Cloud technologies and firm growth rates},
  }
\usepackage{authblk}
\usepackage{comment}
\usepackage[dvipsnames]{xcolor}
\usepackage{float}
\floatstyle{plaintop}
\restylefloat{table}
\usepackage[labelfont=bf,textfont=md]{caption}

\usepackage{setspace}
\usepackage{xcolor}
\newcommand{\rev}[1]{\textcolor{black}{#1}}

\usepackage{natbib} 
\begin{document}
\begin{titlepage}

\title{Scaling Up to the Cloud: \\
	Cloud Technology Use and Growth Rates in Small and Large Firms
  }
\author[1,2]{Bernardo Caldarola}
\author[3,*]{Luca Fontanelli}
\date{\today}
\affil[1]{European Commission, Joint Research Center, Seville, Spain}
\affil[2]{UNU-MERIT and Maastricht University, Maastricht, the Netherlands}
\affil[3]{Università di Brescia, Italy}
\affil[*]{Corresponding author: luca.fontanelli@unibs.it}

\maketitle
\begin{abstract}
\singlespacing

\noindent \rev{Using a unique combination of micro-level data sources on French firms, this paper explores the relationship between cloud technologies and firm growth. Whereas digital technologies have generally been found to disproportionately benefit larger firms, we document that cloud use is positively associated with firm growth, and this association is significantly stronger for smaller firms. Distinguishing cloud services between Infrastructure-as-a-Service (IaaS) and Software-as-a-Service (SaaS), we find that the size-contingent association is driven entirely by SaaS. Crucially, this small firm advantage is conditional on complementary assets as it emerges only among firms endowed with ICT capabilities. Our results are robust across a range of specifications, including instrumental variable strategies. SaaS cloud use is further associated with internal workforce reorganisation toward ICT, managerial, and intermediate occupations and away from manual and clerical ones, particularly among smaller firms, consistent with SaaS cloud operating as an organisational innovation enabling digitalisation and scaling. Finally, we present descriptive sectoral evidence of a negative correlation between cloud and market concentration.}

\end{abstract}
\begin{flushleft}
\textbf{Keywords:}  cloud, ICT, firm growth rate, firm performance, concentration. \linebreak
\textbf{JEL Codes:} L20, L25, O33
\end{flushleft}

\end{titlepage}

\maketitle

	{\footnotesize 
	\section*{\small Acknowledgments} \vspace{-0.3cm} We thank Jordan Bisset, Flavio Calvino, Marco Grazzi, Truls Erikson, Kristina McElheran, Lionel Nesta, Mercedes Teruel, Saleh Zakerinia and participants in the 2024 DRUID, 2024 EARIE conferences, 2024 SBEJ Conference for Young Economists, 2025 TPRI’s Works-in-Progress Seminar Series, 2025 BSE Summer Forum, 2025 ZEW Conference on ICT, 2025 CONCORDi and 2026 IEA conference for insightful comments and discussion. A previous draft of this work has circulated with the title "Cloud technologies, firm growth and industry concentration: Evidence from France". Access to French data benefited from the use of Centre d’accès sécurisé aux données (CASD), which is part of the ``Investissements d’Avenir'' program (reference: ANR-10-EQPX-17) and supported by a public grant overseen by the French National Research Agency. Luca Fontanelli gratefully acknowledges funding from the European Union's Horizon 2020 research and innovation program under the ERC project grant agreement No. 853487 (2D4D). Usual caveats apply. \vspace{.1cm} }

\clearpage\newpage
    
\onehalfspacing
\section{Introduction}

Recent studies have established a positive association between digitalisation and intangible assets on the one hand, and industry concentration on the other \citep[see, among others,][]{bessenindconc2020, Lashkari2024, bajgar2021}. Specifically, the use of information and communication technologies (ICT) disproportionately benefits larger firms \citep{brynjolfssonICTconc2023, babina2021, bessenindconc2020}, owing to scarce complementary assets that larger firms are more likely to possess. However, ICTs may not be alike in this respect. In particular, cloud computing services can substitute for ICT investments \citep{bloomcloud2018, Destefano2023cloud} possibly enabling smaller firms to access otherwise inaccessible IT resources on a utility basis. Yet, to our knowledge, the evidence exploring whether smaller firms derive greater benefits from cloud use than their larger counterparts \rev{-- and, crucially, under which conditions --} remains limited.\footnote{While the relationship between IT outsourcing and firm size is partially addressed in \citet{wangjincloud2024}, the authors are unable to isolate cloud use from IT services in their data and focus primarily on firm age.} 

\rev{Three} arguments motivate this empirical question. First, the costs of ICT adoption and complementary intangible investments are substantial \citep{restatbrynjolfsson2003,deridderaer}, consistent with the well-documented positive relationship between firm size and the use of digital technologies \citep[see][]{zolas2020, calvino2023AI, cirillo, mcelheranNBERAI}. Cloud technologies are therefore particularly valuable for smaller firms, as they reduce the fixed costs associated with digitalisation by allowing firms to store data, operate software, and carry out computationally intensive activities without owning the underlying physical IT infrastructure. Second, notwithstanding the significant decline in cloud service prices during the 2010s \citep{byrne2018cloudprice, coyle2018cloud}, the procurement of cloud services from a market dominated by a small number of large providers \citep[notably Amazon, Microsoft, and Google; see][]{cremersurveycloud} implies the presence of positive markups over the cost of accessing providers' ICT assets. This suggests that when demand for ICT resources is high, as is the case for large firms \citep{Lashkari2024}, purchasing cloud services may not be cost-effective relative to direct ownership. \rev{Three, whether exploiting effectively cloud technologies still requires a base of internal ICT capabilities remains an open question. Two interpretations are plausible a priori. On the one hand, cloud use may allow firms to access an entire ICT system, substituting not only for physical ICT capital \citep{Destefano2023cloud}, but also for other ICT capabilities that they would otherwise need to assemble internally. On the other hand, exploiting cloud effectively may still require a sufficient base of internal ICT capabilities, consistent with empirical evidence about complementarities between ICTs capabilities \citep{tambehitt2012mansci,fontanelliai}.}


In this paper, we examine the heterogeneous impact of the purchase of cloud services on firm size growth rates using a unique combination of four micro-level data sources on French firms: the 2016 and 2018 waves of the French ICT surveys, administrative data from firms' balance sheets, matched employer-employee data, and the French business register. We focus on long-run growth rates, consistent with the view that the effects of digital technologies may take time to materialise owing to the large and complex organisational changes they entail, which are characterised by lengthy implementation lags \citep{restatbrynjolfsson2003, brynjolfsson2018_paradox, acemoglurestreporobot, babina2021, demirer}. We further distinguish between Infrastructure-as-a-Service (IaaS) and Software-as-a-Service (SaaS) cloud technologies. These two modes of use of cloud technologies operate through different channels \citep{Park2023} and lie at opposite ends of the delegation spectrum. IaaS involves outsourcing primarily the underlying IT physical infrastructure, with firms retaining substantial control over their proprietary IT applications and processes, whereas SaaS entails a higher degree of IT delegation, extending to software and associated business functions. Differently from IaaS, SaaS lowers the threshold for accessing advanced software tools and business applications that smaller firms would otherwise need to assemble internally, potentially enabling a broader set of firms to benefit from digitalisation. 

We find that cloud use is positively associated with firm growth, and that this relationship is less pronounced among larger firms. Distinguishing between IaaS and SaaS cloud services, we show that this size-contingent effect is driven entirely by SaaS. 
This result is robust across a range of specifications, including alternative samples and variable definitions, non-linear specifications with employment, sales and age classes, varying growth rate measurement windows, and long-difference estimation. We further address endogeneity concerns through instrumental variable (IV) strategies, including control function approaches and two-stage least squares (TSLS) models designed to account for treatment heterogeneity \citep{wooldridgeCF}. Our instrument exploits spatial variation in lightning strike density at the municipality level, which is associated with investments in IT infrastructure \citep[see also][]{Andersen2012, manacorda, Guriev2021, Goldberg2022, Caldarola2023}. Cloud use requires access to stable and fast internet connections \citep{nicoletti2020, garrison, ohnemus, Destefano2023cloud}, yet lightning strikes generate energy surges that increase infrastructure maintenance costs and slow broadband diffusion \citep{Andersen2012}, while also substantially increasing the frequency of broadband network failures during thunderstorms in the absence of costly mitigation equipment \citep{Schulman2011}. The instrument thus captures the trade-off faced by internet providers between the costs of network expansion and the returns to doing so.

\rev{Furthermore, we explore the conditions under which SaaS cloud technologies are observed to shift the growth constraints of small firms. We find that the SaaS-driven growth advantage of smaller firms is concentrated among those endowed with ICT capabilities, measured by the use of other digital technologies. This supports an ICT complementarity interpretation: SaaS broadens access to digital inputs, but the disproportionate returns accruing to smaller firms are conditional on the presence of complementary ICT assets, which neither IaaS nor SaaS cloud technologies can substitute for}. 

\rev{Next, we expand on evidence that IT use can reshape firms'} workforce composition by producing occupational upgrading towards non-routine cognitive tasks \citep{autor2003skillbiased} and broader upskilling across the workforce \citep{bresnahan2002information}. Growing firms tend to add upper layers to their organisational hierarchy \citep{caliendo2015organization}, and ICTs improve workers' autonomy \citep{bloom2014mansci}, both consistent with an expansion of upper-layer workforce shares among cloud adopters. Finally, if SaaS cloud facilitates firms' digitalisation primarily by substituting physical IT investments \citep{Destefano2023cloud}, then we should also expect the IT component of human capital to increase with SaaS cloud use. We document that SaaS use in smaller firms is associated with an internal reallocation of the workforce toward ICT occupations and upper-layer managerial and intermediate roles, and away from routine manual and clerical tasks. This pattern supports the interpretation of SaaS as an organisational innovation rather than a mere IT cost reduction device.

Finally, we extrapolate from our main findings to provide suggestive evidence on whether cloud use is linked to changes in industrial concentration \citep{brynjolfssonICTconc2023, bessenindconc2020}. Aggregating cloud intensity and market concentration at the two-digit sectoral level, we document a negative correlation between the two. While this association is descriptive in nature, it is consistent with the interpretation that the size-contingent returns to cloud use documented \rev{for smaller, ICT-capable firms} may, in aggregate, be associated with less dispersed market structures across French industries.

\textbf{Related literature and contribution.} Our findings contribute to the growing body of research on the role of cloud technologies in firms' growth and on the mechanisms through which this process unfolds. The extant literature has largely examined the relationship between cloud technologies and firm performance through the lens of firm age rather than size. \citet{Destefano2023cloud} finds that cloud use boosts the growth of young firms, attributing this effect to a reduction in IT investment per employee. \citet{wangjincloud2024} similarly show that young firms disproportionately benefit from IT outsourcing, pointing to mechanisms that mitigate the costs of uncertainty under irreversible investment. More broadly, cloud use is associated with substantial productivity and sales gains \citep{gal,dusocloud2022}. These increase over time \citep{jinbai2022cloud} suggesting a possible role of organisational learning and complementary assets accumulation, consistent with evidence on the dynamics of efficiency in the use of virtual machines \citep{demirer}. The role of cloud in lowering the costs of experimentation is highlighted by \citet{ewens2018cost}, who document that the declining costs of starting new businesses attributable to the advent of cloud infrastructure such as Amazon Web Services have reshaped the venture capital landscape by enabling startups to experiment at lower capital commitment. \citet{fazli2018effects} show that autoscaling features in cloud computing help mitigate price competition by reducing firms' uncertainty about demand fluctuations, thereby improving resource allocation efficiency. Finally, \citet{Park2023} show that SaaS and IaaS operate through distinct channels: SaaS facilitates energy-efficient production processes, whereas IaaS primarily serves to reduce the energy consumption associated with firms' internal IT equipment and infrastructure.

Our findings also add nuance to the literature on ICT and firm growth, which generally suggests that larger firms benefit more from ICT diffusion. Looking specifically at the heterogeneous effect of ICTs across smaller and larger firms, \citet{brynjolfssonICTconc2023} finds that the impact of ICTs on firm size is more pronounced among larger firms, supporting the view that these technologies allow them to replicate best practices across additional production units. Likewise, \citet{babina2021} show that larger firms have ramped up their investments in AI over the past decade, facilitating growth and product innovation -- a rationale that aligns with the findings of \citet{aghionfalling}. Finally, \citet{bessenindconc2020} shows that in IT-intensive industries, the largest firms experienced faster sales growth, and links the diffusion of IT proprietary assets with the increases in industry concentration. \citet{Lashkari2024} document that larger French firms invest a higher share of their sales in IT.

\rev{We contribute to the literature by providing evidence in support of a conditional \textquote{cloud exception} to the broader pattern of ICT-driven growth and concentration. Not all ICTs and cloud technologies affect firms in the same way. In our analysis only SaaS generates a growth premium that is disproportionately large for smaller firms, and this size-contingent effect emerges only among firms with complementary ICT capabilities. Our results therefore sharpen the distinction between ICT access and effective use by differentiating cloud technologies according to the degree of delegation to the cloud provider and by examining their complementarities with firms' digital capabilities. Although SaaS delegates much of the provision and maintenance of software to the cloud provider, its effective use still requires firms to integrate standardised applications with other ICT technologies and adapt business processes accordingly. The growth benefits of SaaS are consequently concentrated among firms that possess these digital capabilities, consistent with evidence on complementarities across digital technologies more broadly \citep{tambehitt2012mansci, mcelheranNBERAI, calvino2023AI, fontanelliai}. The implication is that SaaS can relax the constraint imposed by technology access, while shifting the binding constraint on firm growth toward internal digital readiness.}

\rev{We also document that SaaS use is linked to a reorganisation of the workforce toward digital, intermediate, and managerial functions, particularly among smaller firms. This pattern suggests that cloud operates not
only through the reduction in physical IT investments documented in the existing literature
\citep{Destefano2023cloud}, but also as an organisational innovation, complementary to other ICT capabilities, that enables scaling and digitalisation. At the same time, the expansion of upper-layer workforce shares among SaaS adopters is consistent with the literature on skill-biased technical change \citep{autor2003skillbiased, bresnahan2002information}, while its concentration among smaller firms resonates with work on the role of ICT \citep{coad2024scale} and upper-layer workers in the organisational hierarchy of growing firms \citep{caliendo2015organization}, as well as evidence on the effects of communication technologies on worker autonomy \citep{bloom2014mansci}.}

A key conceptual contribution of our work is to shift the focus from firm age to firm size. While age and size share some common ground, they reflect distinct underlying mechanisms and policy implications. Both young and small firms may lack the complementary assets required for effective digitalisation \citep{brynjolfsson2018_paradox, bresnahan2002information}, but the reasons may differ. Young firms have had limited time to accumulate such assets, and age additionally captures uncertainty about future prospects, which amplifies the risk associated with large upfront and irreversible IT investments \citep{wangjincloud2024}. Size, by contrast, is more directly linked to resource availability such as managerial/IT capabilities \citep{bloomsadunvanreenen2012} and broader scalability constraints that impede growth \citep{garicanolelargereenen2016}, relating directly to the boundary conditions of the firm.

\textbf{Structure of the paper.} The remainder of the paper is organised as follows. Section \ref{sec:data} discusses the sources of data used for the analysis and reports key summary statistics. Section \ref{sec:methods} describes the econometric framework and identification strategy applied in Section \ref{sec:results}, where the main results of the analysis are reported.  Section \ref{sec:mechanism} investigates the mechanisms underlying the heterogeneous effect of SaaS cloud on firm growth. In Section \ref{sec:concentration} we estimate the relationship between cloud use and industry concentration. Section \ref{sec:conclusion} summarises the key findings and discusses policy implications and possible avenues for future research.

\section{Data}
\label{sec:data}

In this section, we discuss the data employed in the analysis and present key summary statistics. Our analysis is based on four sources of microdata.\footnote{The analysis relies on confidential French firm-level data accessed through the Centre d’Accès Sécurisé aux Données (CASD) and commonly used in research \citep[see e.g.][]{domini2021,domini2022,fontanelliai,aghionai} These data are proprietary and subject to strict confidentiality agreements.} First, we use the 2016 and 2018 versions of the French ICT survey (\textit{Enquête sur les Technologies de l'Information et de la Communication (TIC)}), which is managed by the INSEE (the French statistical office).\footnote{Further information about each ICT survey can be found here for 2016 \href{https://www.insee.fr/fr/metadonnees/source/operation/s1062/presentation}{https://www.insee.fr/fr/metadonnees/source/operation/s1062/presentation} and here for 2018 \href{https://www.insee.fr/fr/metadonnees/source/operation/s1391/presentation}{https://www.insee.fr/fr/metadonnees/source/operation/s1391/presentation}.} Each wave of the survey includes a rotating sample counting approximately 9000 firms from both manufacturing and non-financial market-services sectors. The sample is representative for firms with 10 or more employees and is exhaustive for those with over 500 employees.\footnote{It is therefore challenging to exploit the panel dimension of these datasets. Approximately sixteen hundred firms are present in both the 2016 and 2018 ICT surveys, and are mostly large.}  We exclude firms located in Corsica and overseas provinces. The survey questions on cloud technologies focus on the purchase of cloud services in 2016 and 2018.\footnote{The survey is distributed in the early months of the reference year. The questions about advanced digital technologies are updated annually, and the ICT surveys run in different years may not include questions about the same technologies. Additional questions about cloud technologies are present in the 2020 and 2021 waves. However, in the context of our identification strategy (see Section \ref{sec:methods}), the use of this waves implies the inclusion of the COVID-19 pandemic years in the dataset. The pivotal role of digital technologies during the pandemic makes it challenging to precisely estimate the effect of cloud on performance in normal times. We chose to employ the 2016 and 2018 ICT surveys accordingly. We provide a robustness check confirming our results when using the 2020 and 2021 versions of the ICT survey in the Appendix (see Table \ref{tab:robsample}, Column 1).} Additionally, the ICT survey can be merged to other sources of French firms' data thanks to the \emph{Siren} code, a unique identifier attributed to  French companies at their birth. Part of the ICT survey is dedicated to questions on cloud use by firms. Specifically, firms are asked the following question: \textquote{Does your enterprise buy cloud computing services? (Excluding cloud services provided for free)}. In the survey, cloud services are defined as follows:

\begin{displayquote}
Cloud computing (or cloud) refers to computing services used over the internet to access software, computing power, storage capacity, etc. These services must have the following characteristics:
\begin{itemize}
\item They are delivered by servers from service providers.
\item They are easily scalable up or down (for example, the number of users or changes in storage capacity).
\item Once installed, they can be used \textquote{on-demand}, without human interaction with the provider.
\item They are paid either by the user or based on the capacity used or services provided.
\end{itemize}
Cloud computing may include connections via a virtual private network (VPN).
\end{displayquote}

Furthermore, the survey provides information on the different types of cloud services purchased by firms, distinguishing them into seven non-exclusive categories: mail, data storage, file storage, accounting software, office software, customer relationship management (CRM) software, and computing power. We define a cloud user as a firm that purchases cloud services in at least one of the latter six categories. We discard the first category of cloud usage (mail), as it is unlikely to produce organisational changes in the firm's structure and, therefore, may not capture the effects of cloud on firm performance. Our main cloud use variable measures whether firms use cloud technologies or not.  

Additionally, we provide results for different categories of cloud usage. Following the standard taxonomy of cloud services \citep{Park2023}, we regroup the remaining categories into two broader types: cloud as infrastructure (IaaS) and cloud as software (SaaS).\footnote{The ICT survey does not include items that map cleanly onto Platform-as-a-Service (PaaS) -- an intermediate in the spectrum covered by IaaS and SaaS -- so we restrict our classification to IaaS and SaaS.} A firm is considered to use IaaS when it purchases cloud services for storing data, storing files, or borrowing external IT processing capacity, and to use SaaS when it purchases cloud services for accounting, office, or CRM software. In consequence of these choices, our cloud variable captures only purchased cloud services, excluding free and email services, and reflects use patterns observed in 2016 and 2018. Several considerations bear on the external validity of our findings. First, our measure captures whether a firm purchased cloud services in the survey year, but does not record the intensity of use. Second, the exclusion of free-tier services means that our measure captures a deliberate and costly commitment to cloud infrastructure, which may be more informative about organisational intent than broader use indicators, but also means that our results may not generalise to settings where free cloud access is prevalent. Third, our findings are technologically contingent on the digital tools available during the period under study, which overlap with the early diffusion of AI technologies, but predates the one of generative AI tools.

The ICT survey also collects information on firms’ use of other digital technologies. Specifically, in both the 2016 and 2018 waves, firms are asked about their engagement in e-commerce activities and their use of big data analytics. Moreover, the survey includes questions on broadband connectivity, specifying whether the firm has no broadband access or a connection speed of less than 2, between 2 and 10, between 10 and 30, between 30 and 100, or 100 Mbit/s and above. 

Second, we match the ICT survey with the administrative data from French firms' balance sheets (FARE) covering the 2011-2018 period.\footnote{Additional details about this dataset can be accessed here: \url{https://www.casd.eu/en/source/annual-structural-statistics-of-companies-from-the-esane-scheme/}.} This dataset provides information on firm sales, age, employment, geographical location, exporter status, and physical or intangible capital.\footnote{Data on sales and capital are provided in real terms. Sales and physical capital have been deflated at the 2-digit sector level. Data on intangible capital have been deflated exploiting the deflators provided by INTANPRO-EUKLEMS \citep{bontadini2023euklems}.} Intangible capital is not available before 2009.\footnote{However, in Section \ref{subsec:robustness} we discuss robustness checks spanning until the year 2008.} These variables allow us to provide a complete picture of firms buying cloud technologies, and to control for potential links between size growth, firm, and geographical characteristics. Third, we employ the information on the stocks of establishments by firm from the French business register. This data is used to build a binary variable indicating if the firm is multi establishment.

Finally, we match the ICT survey with French employer-employee data (DADS) in 2011-2018.\footnote{Further information about DADS here \url{https://www.casd.eu/en/source/all-employees-databases-business-data}.} This data allow us to build the firm-level share of hours worked by ICT technical workers and by ones specialised in R\&D hired by the firm (named ICT share and R\&D share hereafter), and the average hourly wage of managers and engineers in a firm.\footnote{It is worth noting that ICT workers are part of the techies definition used in \citet{harrigan}. The techies definition encompasses all occupations within the 2-digit classes 38 (executives and engineers) and 47 (Technicians) of the 2003 French PCS classification. The mentioned PCS codes cover roles such as R\&D personnel in IT, computer engineers, developers, database administrators, and IT technicians. Further details and information on the PCS classification can be found here \url{https://www.insee.fr/fr/statistiques/fichier/2401328/Brochure_PCS_ESE_2003.pdf}.} We consider ICT technical workers to be employees falling within the 4-digit classes 388a, 388b, 388c, 388d, 388e, 478a, 478b, 478c, 478d, and 544a of the 2003 French PCS classification (\textit{Nomenclature des professions et catégories socioprofessionnelles}). Instead R\&D technical workers fall within the 4-digit classes 383a, 384a, 385a, 386a, 388a, 473a, 473b, 474a, 475a, and 478a, as suggested by the classification of occupations into functions provided by the French National Statistical Institute.\footnote{The classification can be found here \url{https://www.insee.fr/fr/statistiques/1893116}.} 
These classes specifically target occupations with a significant focus on ICTs and R\&D.

\subsection{Summary statistics}
\label{subsec:summ_stats}

Before investigating the relationship between cloud use and firm growth rates, we sketch out the characteristics of our sample of firms, highlighting some general differences between cloud users and non-users. To start with, the upper block in Table \ref{tab:users} shows the share of cloud-adopting French firms grew by a remarkable 23.63 percent (6.39 percentage points) between 2016 and 2018, indicating a rapid diffusion of cloud technologies. Foreseeably, the share of cloud users increased in each sector considered. Cloud technologies are most frequently adopted by firms in the ICT sector: their use rates span from 50.7 percent of ICT firms in 2016 to 59.96 percent in 2018. Large changes in the rate of use are registered across all sectors, suggesting that cloud technologies are rapidly diffusing everywhere. In 2018, professional and scientific, real estate, and administrative firms displayed large rate of cloud use with respect to other sectors. Low-productivity sectors such as accommodation, transportation services, and wholesale and retail displayed lower levels of use, with limited growth over time. The most represented industry in the sample, manufacturing, also experienced sizable growth in the use of cloud technologies, moving from 27.59 percent of adopters in 2016 to 33.95 percent in 2018 -- a change of 6.36 percent, corresponding to a growth rate of 23.05 percent.

We now describe the general characteristics of cloud users versus non-users. As shown in Table \ref{tab:averages}, among firms using cloud, IaaS was the most commonly used service in both years (around 90\% of cloud users), although SaaS cloud has been growing more rapidly than IaaS between 2016 and 2018 (from 67.17 percent of firms in 2016 to 77.2 percent in 2018). Within IaaS cloud, the fastest growing (and less common) type is cloud for computing power -- which grew from 22.79 percent of firms in 2016 to 26.76 percent in 2018. The vast majority of firms had already adopted cloud services for storage -- 89.3 percent in 2016 and 90.64 percent in 2018, likely because these services were the first to be commercialised, and have reached earlier maturity.  Overall, Table \ref{tab:averages} also shows that cloud-adopting firms tend to be older and remarkably larger in terms of sales. They also own a bigger stock of physical and intangible capital, employ a higher share of ICT workers (more than twice as bigger than non-adopters), and are more likely to export. Finally, cloud adopting firms are also more likely to deploy more than one productive plant or unit. 

\bigskip

\begin{table}[!htbp]
    \centering \scalebox{.6}{
        \begin{tabular}{lcc}
            \toprule \addlinespace
            \textbf{France} & 2016 & 2018 \\\addlinespace\cmidrule(lr){2-3}\addlinespace
            All Firms   & 27.04\% & 33.43\% \\ \addlinespace \midrule\addlinespace
            \textbf{Industry} & 2016 & 2018 \\\addlinespace\cmidrule(lr){2-3}\addlinespace
            Accomodation \& Food   & 16.42\% & 19.66\% \\\addlinespace
            Administrative  & 28.91\% & 35.23\% \\\addlinespace
            ICT  & 50.70\% & 59.96\% \\\addlinespace
            Manufacturing   & 27.59\% & 33.95\% \\\addlinespace
            Professional \& Scientific & 38.70\% & 43.09\% \\\addlinespace
            Real Estate   & 29.71\% & 43.24\% \\\addlinespace
            Transportation \& Storage  & 23.29\% & 29.88\% \\\addlinespace
            Utilities \& Construction  & 18.45\% & 24.63\% \\\addlinespace
            Wholesale \& Retail    & 23.02\% & 29.75\% \\\addlinespace\bottomrule
    \end{tabular}}
    \caption{Share of cloud users in 2016 and 2018: total for France and by industry.}
    \label{tab:users} 
\end{table}

\newpage

\begin{table}[!htbp]
    \centering \scalebox{.65}{
        \begin{tabular}{lcccc}
        \toprule
        Year &\multicolumn{2}{c}{2016}&\multicolumn{2}{c}{2018}\\  \cmidrule(lr){2-3} \cmidrule(lr){4-5}
        Cloud Use                              & No        & Yes        & No        & Yes        \\
        \midrule \addlinespace
        Cloud -- IaaS                           &          & 89,30\%  &          & 90,64\%  \\\addlinespace
        Cloud -- IaaS (Storage)                        &          & 87,56\%  &          & 88,98\%  \\\addlinespace
        Cloud -- IaaS (Computing Power)                &          & 22,79\%  &          & 26,76\%  \\\addlinespace 
        Cloud SaaS -- (Software)                       &          & 67,17\%  &          & 77,20\%  \\\addlinespace \midrule \addlinespace
        Age                                    & 28,19    & 31,33    & 28,91    & 32,57    \\\addlinespace
        Sales (Thousands \euro)                & 85552,46 & 317207,20& 67354,86 & 339490,90\\\addlinespace
        Employment               & 222,94   & 986,71   & 181,63   & 940,96   \\\addlinespace
        Physical Capital (Thousands \euro)     & 34611,82 & 183282,40& 21437,02 & 194226,80\\\addlinespace
        Intangible Capital (Thousands \euro)   & 355,94   & 2087,29  & 183,41   & 2009,36  \\\addlinespace
        ICT Share                              & 2,91\%   & 8,90\%   & 2,59\%   & 8,48\%   \\\addlinespace
        R\&D Share                             & 2,95\%   & 7,54\%   & 2,92\%   & 6,95\%   \\\addlinespace
        Exporter                               & 42,61\%  & 63,60\%  & 40,51\%  & 63,99\%  \\\addlinespace
        Multi Establishment                    & 42,28\%  & 65,74\%  & 41,12\%  & 64,97\%  \\\addlinespace
        Hourly Wage (Managers \& Engineers, €)  & 26.78    &	32.11   & 25.35    & 33.67 \\\addlinespace\bottomrule
        \end{tabular}}
    \caption{Summary statistics by cloud user and year}
    \label{tab:averages}
\end{table}

\section{Methods}
\label{sec:methods}

In this paper, we study the relationship between cloud purchases and firms' sales growth. Specifically, our baseline specification uses the 5-year logarithmic difference in sales as the dependent variable. The baseline regression model reads as follows:

\begin{equation}\label{eq:baseline_OLS}
\begin{split}
  & \text{Sales Growth}_{i, t, t-5} = \\
  & a + \beta_1 \text{Cloud}_{i,t} + \beta_2 \text{Cloud}_{i,t} \times 
    \text{Log-Sales}_{i,t-5} + \beta_3\text{Log-Sales}_{i,t-5} + 
    \mathbf{\beta_X}\mathbf{X}_{i,t-5} + \\
  & + \text{2-digit Ind.}_{j} + \text{Region}_{r} + \text{Year}_{t} + 
    \epsilon_{i,t}
\end{split}
\end{equation}

where $\text{Sales Growth}_{i, t, t-5}$ is the logarithmic difference between sales in year $t$ and year $t-5$, and $\text{Cloud}_{i,t}$ is a dummy variable taking value 1 if the firm reported purchasing cloud services in year $t$, or alternatively the vector of IaaS and SaaS dummies. Following the empirical strategy of \citet{formanKM}, controls in $\mathbf{X}_{i,t-5}$ are measured at the beginning of the period, in $t-5$, to mitigate reverse causality concerns. These include the logarithms of age, physical capital, and intangible capital, the average hourly wage of managers and engineers, the shares of hours worked in ICT and R\&D occupations, and dummies for exporter and multi-establishment status. $\text{Log-Sales}_{i,t-5}$ is included separately rather than absorbed into $\mathbf{X}_{i,t-5}$, given its role in the interaction term: $\beta_3$ captures the baseline relationship between initial size and subsequent growth, while $\beta_2$ isolates the extent to which the 
cloud-growth relationship varies with firm size. Fixed effects for 2-digit industries ($j$), regions ($r$), and years ($t$) are included throughout. 

\textbf{Long-run growth rates.} We test the relationship between cloud use and growth using long-term growth rates. Doing otherwise (that is, employing short-term growth rates such as annual growth rates as the dependent variable) may fail to capture the relationship of interest for two key reasons. First, extensive literature shows that short-term firm growth rates often align with Gibrat’s Law \citep{gibrat}, particularly in the case of less young surviving firms \citep{santarelli2006,lotti2003,lotti2009,fontanellisurvey}.\footnote{According to Gibrat’s model, firm size evolves as $\log s_{i,t} = \log s_{i,0} + \sum_{\tau=1}^{t} \epsilon_{\tau}$, where the growth rate $\epsilon_{\tau}$ is an identically and independently distributed random variable.} This implies that short-term fluctuations are largely stochastic and may not capture growth drivers materialising in the long-run.
Second, as highlighted in several studies \citep{restatbrynjolfsson2003,tambehitt2012mansci,brynjolfsson2018_paradox,acemoglurestreporobot}, the effects of the diffusion of digital technologies often require time to materialise. This delay stems from the uncertainties and implementation lags caused by the substantial and complex organisational changes associated with ICT adoption. This reasoning applies to cloud-based use of virtual machines as well \citep{demirer}. The use of long-run changes, including firm growth rates, is common in the literature studying the effects of digital technologies \citep{babina2021, forman2012,formanKM}.

\textbf{Measuring cloud use in $t$.} While the ICT surveys do not provide information on the year of adoption and first use of cloud by firms, they do indicate whether a cloud service was purchased by the firm in the survey year. The unavailability of information on the first year of cloud use by firms introduces a measurement challenge. Specifically, the adoption of cloud captured by $Cloud_{i,t}$ could have happened at any point before time $t$, but not after that. This maximises the likelihood that the first instance of cloud adoption will have occurred between the beginning and the end of the long term growth period.

Consistently with the empirical strategies adopted in the literature measuring the effects of ICT \citep{forman2012,formanKM}, we measure cloud use in $t$. The use of cloud services by firms before 2009 was highly unlikely even in a digitally advanced country such as the US \citep{bloomcloud2018}, due to the high prices associated with cloud service provision until the early 2010s \citep{byrne2018cloudprice, coyle2018cloud}. This suggests that French firms in our sample, observed in 2016 and 2018, likely began adopting cloud technologies in the early 2010s. Therefore, measuring cloud in $t$ better captures these early adoption phases, and thus picks up the lion's share of performance gains owing to cloud use. Furthermore, measuring cloud use at $t-5$ implies the use of post-2019 data risks, conflating the effects of cloud use with pandemic-induced dynamics, as COVID-19 not only accelerated digital adoption \citep{avalos,calvinocovid} but also potentially influenced sales disparities between users and non-users through resilience mechanisms associated with broader digitalisation rather than cloud use per se.

Figure \ref{fig:vol_window}, adapted from \citet{fontanellivolai}, helps navigate the measurement issue, illustrating the advantages and disadvantages of this procedure in five different cases in the context of the 2018 ICT survey. Firms 1 and 2 are correctly classified. Firms 3, 4 and 5 introduce measurement errors. Firm 3 only used cloud for a few periods. Firm 4, despite having used cloud, is classified as a non-user. Firm 5 starts using cloud after the year of observation, and is therefore also classified as a non-user in 2018. First, if cloud use has a positive effect on growth, our approach appears conservative as a first approximation. By extending firm status over the period, we introduce a potential bias that underestimates growth for cloud users (Firm 3) and overestimates growth for non-cloud users (Firm 4). The case of Firm 5 illustrates that the cloud–growth relationship may also be underestimated when using forward-looking growth rates. Similarly, cases such as Firm 4 may distort the estimated relationship between forward growth and cloud adoption.

This bias makes it more challenging to detect growth differences between the two groups, meaning that any significant difference we do observe is likely a lower bound of the true effect. 

\begin{figure}[!ht]
  \centering
  \caption{Computing sales growth rates.}
    \label{fig:vol_window}
  \begin{minipage}{0.9\linewidth}
    \centering
        \begin{tikzpicture}
      \tikzset{
        uniform cell/.style={
          draw,
          align=center,
          text width=0.9cm,        
          minimum height=0.8cm,       
          inner sep=2pt,
          anchor=center,
          execute at begin node=\strut  
        }
      }
  
      \matrix (m) [matrix of nodes,
        nodes in empty cells,
        nodes={uniform cell},
        column sep=-\pgflinewidth,   
        row sep=-\pgflinewidth,
        outer sep=0pt
      ] {
        0 (0)  & 0 (0) & 0 (0)   & 0 (0)   & 0 (0)   & 0 (0)   & {\bf 0 }  & 0 (0) \\
         0 (0)  & 1 (1)  & 1 (1)   & 1 (1)   & 1 (1)   & 1 (1)   & {\bf 1 }  & 1 (1) \\
         \textcolor{red}{1 (0)}  & \textcolor{red}{1 (0)}   & \textcolor{red}{1 (0)}   & \textcolor{red}{1 (0)}  & 1 (1)   & 1 (1)   & {\bf 1 }  & 1 (1) \\
       \textcolor{red}{0 (1)}  & \textcolor{red}{0 (1)}   & \textcolor{red}{0 (1)}   & \textcolor{red}{0 (1)}   & \textcolor{red}{0 (1)}   & 0 (0)   & {\bf 0 }  &  0 (0) \\
       0 (0)  & 0 (0) & 0 (0)   & 0 (0)   & 0 (0)   & 0 (0)   & {\bf 0 }  &  \textcolor{red}{0 (1)} \\
      };

      \node[left=5pt] at (m-1-1.west) {Firm 1};
      \node[left=5pt] at (m-2-1.west) {Firm 2};
      \node[left=5pt] at (m-3-1.west) {Firm 3};
      \node[left=5pt] at (m-4-1.west) {Firm 4};
      \node[left=5pt] at (m-5-1.west) {Firm 5};
      
      \foreach \i [count=\j from 1] in {{\bf 2012}, 2013, 2014, 2015, 2016, 2017, {\bf 2018}, 2019} {
        \node[above=5pt] at (m-1-\j.north) {\i};
      }
      
    \draw[ForestGreen, thick, dashed] ($(m-1-2.north west)+(0.1,-0.1)$) rectangle ($(m-5-7.south east)+(-0.1,0.1)$);
      
      \draw[thick] (m-1-1.north west) rectangle (m-4-8.south east);
      
    \end{tikzpicture}

    \vspace{-1.0em}\begin{quote} {\footnotesize \textit{Notes:} Cloud
        \textquote{user} (1) and \textquote{non-user} (0) status over 2012-2019 for four
        hypothetical firms. The 2018 ICT survey provides factual information on
        their status of users in 2018. Since in other years this
        information is not available, we propagate it backward to cover the 5-years time span 2013-2018 (dashed green border). Firms are assumed to have at most one switch from \textquote{non-user} to \textquote{user}. To illustrate
        classification mistakes, we report in parenthesis the
        hypothetical true status. If the two indicators within a cell
        match, our procedure correctly classifies the firm as a user
        or non-user (cells with black numbers); if they do not match,
        the classification contains some measurement error (cells with red numbers).
        \par} \end{quote}
  \end{minipage}
\end{figure}

\textbf{Control variables.} The vector of variables $\text{X}_{i,t-5}$ includes a comprehensive set of time-varying firm characteristics. The logarithm of age, sales, physical capital, and intangible capital; the share of hours worked in ICT and R\&D occupations; the logarithm of the average hourly wage of managers; dummies for export and multi establishment status; and fixed effects for 2-digit industries, regions, and years. The ICT and R\&D shares serve as a proxy for the intensity of digitalisation and innovation within the firm and, in our regression setting, clean the relationship between cloud and performance from the correlation between cloud and other performance-enhancing innovations and digital technologies. The average hourly wage of managers is an approximate measure of the quality of managers in the firm.

The inclusion of firms' sales among the control variables mitigates a key source of bias. Larger firms are more likely to innovate and adopt digital technologies, including cloud. Therefore, controlling for sales reduces the potential confounding effects of other digital technologies and innovations adopted by larger firms. Firm age is also included as a control to account for new managerial and ICT capabilities potentially affecting both cloud use and performance \citep{bloomcloud2018}. Since younger firms may be more likely to adopt emerging technologies like AI \citep{Calvino2026respol}, controlling for age helps reduce bias from omitted variables while also cleaning the size-cloud interaction.

Controlling for firm capital further addresses endogeneity issues. Intangible capital, for instance, includes the firm-level value of complementary digital technologies, such as proprietary data and software, and factors which may impact firm growth rates, as it also includes the value of assets such as patents and trademarks \citep{corrado2021}. Physical capital affects the feasibility of cloud use, as firms with lighter capital structures may find cloud technologies more advantageous. 
Export and multi establishment dummies are included to control for firms’ access to multiple markets, addressing the self-selection of firms with higher growth potential and growth strategies prioritising growth into cloud usage. Finally, we include fixed effects for 2-digit NACE industries, regions, and years.

\section{Results}
\label{sec:results}

\begin{table}[!htbp] \centering 
    \begin{threeparttable}
        \scalebox{0.65}{
            \begin{minipage}{\textwidth}
                 \begin{tabular}{lcccccc}
                    \toprule
                    & (1)        & (2)        & (3)        & (4)        & (5)        & (6)        \\
                    \addlinespace\cmidrule(lr){2-7}
                    Cloud (t)                               & 0.388***   & 0.302***   & 0.101***   & 0.267***   & 0.265***   & 0.294***   \\
                    & (0.084)    & (0.083)    & (0.018)    & (0.076)    & (0.069)    & (0.075)    \\\addlinespace
                    Log Sales (t-5)                         & -0.034***  & -0.044***  & -0.047***  & -0.041***  & -0.043***  & -0.045***  \\
                    & (0.005)    & (0.004)    & (0.006)    & (0.006)    & (0.006)    & (0.006)    \\\addlinespace
                    Cloud (t)$\times$Log Sales (t-5)        & -0.025***  & -0.019***  &            & -0.017***  & -0.018***  & -0.012**   \\
                    & (0.007)    & (0.007)    &            & (0.006)    & (0.006)    & (0.005)    \\\addlinespace
                    Log Age (t-5)                           &            &            & -0.089***  & -0.089***  & -0.089***  & -0.065***  \\
                    &            &            & (0.011)    & (0.011)    & (0.011)    & (0.009)    \\\addlinespace 
                    Share of ICT Hours (t-5)                &            &            & 0.186**    & 0.180**    & 0.181**    & 0.175**    \\
                    &            &            & (0.071)    & (0.071)    & (0.068)    & (0.069)    \\\addlinespace
                    Share of R\&D Hours  (t-5)         &            &            & 0.214***   & 0.215***   & 0.229***   & 0.226***   \\
                    &            &            & (0.052)    & (0.052)    & (0.054)    & (0.052)    \\\addlinespace
                    Log Physical Capital (t-5)              &            &            & -0.003     & -0.003     & 0.010      & -0.003     \\
                    &            &            & (0.009)    & (0.009)    & (0.008)    & (0.008)    \\\addlinespace
                    Log Intangible Capital (t-5)               &            &            & 0.011      & 0.011      & 0.011      & 0.011      \\
                    &            &            & (0.008)    & (0.008)    & (0.008)    & (0.008)    \\\addlinespace
                    Exporter (t-5)                          &            &            & 0.016      & 0.015      & 0.012      & 0.010      \\
                    &            &            & (0.012)    & (0.012)    & (0.012)    & (0.011)    \\\addlinespace
                    Multiplant (t-5)                        &            &            & -0.011     & -0.011     & -0.013     & -0.010     \\
                    &            &            & (0.009)    & (0.009)    & (0.009)    & (0.008)    \\\addlinespace
                    Log Avg Hourly Wage Managers (t-5)      &            &            & 0.008**    & 0.006*     & 0.006*     & 0.005      \\
                    &            &            & (0.003)    & (0.003)    & (0.003)    & (0.003)    \\\addlinespace
                    E-commerce (t)                          &            &            &            &            & 0.074***   & 0.177      \\
                    &            &            &            &            & (0.020)    & (0.125)    \\\addlinespace
                    Big Data (t)                            &            &            &            &            & 0.041***   & 0.100*     \\
                    &            &            &            &            & (0.008)    & (0.057)    \\\addlinespace
                    Cloud (t)$\times$Log Age (t-5)          &            &            &            &            &            & -0.029     \\
                    &            &            &            &            &            & (0.018)    \\\addlinespace
                    E-commerce (t)$\times$Log Sales (t-5)   &            &            &            &            &            & 0.005      \\
                    &            &            &            &            &            & (0.005)    \\\addlinespace
                    E-commerce (t)$\times$Log Age (t-5)     &            &            &            &            &            & -0.051*    \\
                    &            &            &            &            &            & (0.025)    \\\addlinespace
                    Big Data (t)$\times$Log Sales (t-5)     &            &            &            &            &            & 0.000      \\
                    &            &            &            &            &            & (0.006)    \\\addlinespace
                    Big Data (t)$\times$Log Age (t-5)       &            &            &            &            &            & -0.021     \\
                    &            &            &            &            &            & (0.013)    \\\addlinespace
                    \addlinespace\midrule\addlinespace
                    Observations                            & 16,216     & 16,216     & 16,216     & 16,216     & 16,216     & 16,216     \\
                    Adj R2                                  & 0.041      & 0.068      & 0.096      & 0.097      & 0.102      & 0.104      \\
                    Industry, Reg., Year FE                 &            & X          & X          & X          & X          & X          \\
                    \bottomrule
                 \end{tabular}
                \begin{tablenotes}
                    \item \textit{Note}: The table reports the results of the OLS estimation of Equation \ref{eq:baseline_OLS}. The dependent variable is the firm's sales growth rate between $t$ and $t-5$. ICT use binary variables are measured at time $t$ (cloud, e-commerce and big data analytics), while remaining firm-level control variables are measured at $t-5$. Log Sales, Physical, and Intangible Capital are expressed in Log of Euros. Age is the firm age in years since the establishment. ICT and R\&D shares express, respectively, the share of hours worked in ICT and R\&D occupations. Exporter is a dummy that identifies whether the firms is engaged in exporting activities. Multi Establishment identify whether the firm has more than one economic establishment. Log Average Hourly Wage Manag. \& Eng. measures the log of the hourly wage of management and engineers, expressed in Euros. All regressions include 2-digit industry, region and year dummies. Standard errors are clustered at the NACE 2-digit level.  *** p$<$0.01, ** p$<$0.05, * p$<$0.1. 
                \end{tablenotes}
            \end{minipage}
        }
    \end{threeparttable}
    \caption{Baseline model}
    \label{tab:OLS}
\end{table}

\textbf{Cloud use and firm growth rates.} We examine the association between the use of cloud technologies and firm growth in France over the period 2016–2018, as specified in Equation \ref{eq:baseline_OLS}. Results are reported in Table \ref{tab:OLS}. Column (1) shows a positive and statistically significant relationship between cloud use and long-run sales growth rates. The magnitude of this association declines after the inclusion of industry, region, and year fixed effects (Column 2), but remains statistically significant across specifications. The interaction between cloud use and baseline firm size (log sales) bears a negative and significant coefficient, indicating that the association between cloud and growth is weaker for larger firms (Columns 1–2 and 4–6). Consistently, the coefficient on firm size is negative across all specifications, suggesting that larger firms grow more slowly than smaller ones. \rev{The estimated growth premium of cloud users versus non-users decreases monotonically with firm size, ranging from 15.06\% at the 10th percentile to 5.91\% at the 90th percentile of the sales distribution, and standing at 11.7\% at the median (see Table \ref{tab:effects}, Column 2 in the Appendix).}\footnote{\rev{Recall that the dependent variable is defined as the five-year log-difference in sales, so that the coefficients reported in Table \ref{tab:effects} approximate, but do not exactly equal, percentage growth rates.}}

Similarly to firm size, firm age is negatively associated with growth. The shares of ICT and R\&D employment are positively correlated with firm performance, in line with existing evidence \citep{coadquantile,brynjolfssonICTconc2023}. Exporting status, multi-establishment structure, and capital inputs are not significantly associated with growth once we control for firm characteristics and fixed effects. Finally, firms with higher growth rates tend to pay higher wages to their managers.

Columns 5 and 6 in Table \ref{tab:OLS} further investigate whether the interaction between cloud use and firm size is driven by confounding factors. First, while the growth premium from cloud declines with baseline size, firm size is correlated with age, and prior evidence \citep{Destefano2023cloud,wangjincloud2024} suggests that younger firms may benefit more from cloud technologies. Second, the smaller effect estimated for cloud use on the growth of larger firms may reflect decreasing returns to digital investments or demand-side constraints, whereby firms operating in less elastic markets are less able to translate technology use into sales growth.\footnote{However, existing evidence does not strongly support decreasing returns to ICT or AI investments in terms of firm growth \citep{brynjolfssonICTconc2023,babina2021}. Moreover, the view that smaller firms grow faster than larger firms, conditional on survival, has been challenged once age is properly controlled for \citep{coadHGF,haltiwanger2013}.} Finally, given the complementarity among digital technologies \citep{oecd2026_digital_technology_diffusion}, cloud use may proxy for broader digital transformation rather than having an independent association. To address these concerns, we extend Equation \ref{eq:baseline_OLS} by estimating a horse-race specification that includes interactions between multiple digital technologies (cloud, big data analytics, and e-commerce) and firm size and age.

Both e-commerce and big data analytics are positively and significantly associated with firm growth (Column 5, Table \ref{tab:OLS}), in line with existing evidence. E-commerce may enhance performance by expanding market access \citep{couture} and reducing search costs \citep{goldmanis}, while big data analytics can improve productivity through data-driven decision-making and process innovation \citep{andresBDA,contiBDA}. Importantly, the interaction between cloud use and firm size remains negative and statistically significant, although its magnitude is slightly reduced, once introduced the full set of interactions (Column 6, Table \ref{tab:OLS}). In contrast, the interaction between cloud and firm age, as well as the interactions involving other technologies, are not statistically significant, indicating that the moderating role of firm size in the cloud–growth relationship is not driven by firm age or by the use of other digital technologies.

\textbf{Heterogeneity across cloud types.} We further examine how cloud use relates to firm growth by distinguishing between cloud-as-infrastructure (IaaS) and cloud-as-software (SaaS), estimating Equation \ref{eq:baseline_OLS} with two non-mutually exclusive indicators of cloud use. Table \ref{tab:OLS_types} reports the results. While the positive association between cloud use and firm growth is confirmed for both types (Column 1), the interaction coefficients differ markedly across cloud models (Column 2). The only negative and significant interaction with firm size is for SaaS, indicating that the growth association is stronger for smaller firms adopting cloud-based software. By contrast, IaaS displays no significant base coefficient or interaction with size.

These patterns are consistent with the distinct functional roles of each cloud model. SaaS lowers the cost and complexity of accessing advanced software applications and business processes without requiring substantial upfront investment in physical IT infrastructure, disproportionately benefiting smaller firms that are less likely to possess established in-house IT systems. IaaS, by contrast, provides raw computing and storage resources whose effective use depends heavily on proprietary software that may be limited precisely among small firms \citep{bessenindconc2020}. Absent these inputs, IaaS adoption may not yield immediate performance gains, and may instead involve implementation lags and additional investments \citep{brynjolfsson2018_paradox, brynjolfsson2021}, consistent with evidence of efficiency in the use of virtual machines and its dynamics, with slow learning curves and time-persistent differentials \citep{demirer}. \rev{We return in Section \ref{sec:mechanism} to the conditions under which this small-firm advantage materialises.}

\begin{table}[!ht] \centering 
    \begin{threeparttable}
        \scalebox{0.65}{
            \begin{minipage}{\textwidth}
                 \begin{tabular}{lcc}
                    \toprule
                    & (1)        & (2)        \\
                    \addlinespace\cmidrule(lr){2-3}
                    Cloud IaaS (t)                          & 0.073***   & 0.090      \\
                    & (0.013)    & (0.066)    \\\addlinespace
                    Cloud SaaS (t)                          & 0.045**    & 0.313***   \\
                    & (0.021)    & (0.079)    \\\addlinespace
                    Cloud IaaS (t)$\times$Log Sales (t-5)   &            & -0.002     \\
                    &            & (0.006)    \\\addlinespace
                    Cloud SaaS (t)$\times$Log Sales (t-5)   &            & -0.026***  \\
                    &            & (0.007)    \\\addlinespace
                    Log Sales (t-5)                         & -0.047***  & -0.040***  \\
                    & (0.006)    & (0.006)    \\\addlinespace
                    \addlinespace\midrule\addlinespace
                    Observations                            & 16,216     & 16,216     \\
                    Adj R2                                  & 0.096      & 0.099      \\
                    Industry, Reg., Year FE                 & X          & X          \\
                    \bottomrule
                 \end{tabular}
                \begin{tablenotes}
                    \item \textit{Note}: The table reports the results of the OLS estimation of Equation \ref{eq:baseline_OLS}. The dependent variable is the firm's sales growth rate between $t$ and $t-5$. Cloud use variables are measured at time $t$ (IaaS and SaaS cloud), while remaining firm-level control variables are measured at $t-5$. Firm controls are measured in $t-5$ and include the logarithms of age, tangible and intangible capital, the share of workers specialised in ICT and R\&D roles, the average hourly wage of managers and engineers, and two dummies for exporter and multi establishment status. All regressions include 2-digit industry, region and year dummies. Standard errors are clustered at the NACE 2-digit level.  The complete version of this table, including additional coefficients of controls, is available upon request. *** p$<$0.01, ** p$<$0.05, * p$<$0.1. 
                \end{tablenotes}
            \end{minipage}
        }
    \end{threeparttable}
    \caption{Cloud types}
    \label{tab:OLS_types}
\end{table}

\textbf{Heterogeneity across sectors.} Finally, we examine heterogeneity across sectors by estimating Equation \ref{eq:baseline_OLS} separately for manufacturing, advanced business services,, other services and utilities and construction (Table \ref{tab:sectors} in Appendix). The positive association between SaaS cloud and firm growth, and its size-contingent nature, is confirmed in manufacturing (Column 1) and other services (Column 2), where both the SaaS coefficient and its interaction with size are statistically significant and carry the expected signs. By contrast, no significant associations are detected for advanced business services (Column 3) or utilities and construction (Column 4). This pattern is broadly consistent with the idea that SaaS cloud confers the greatest benefits in sectors where firms are less digitalised, but where digitalisation remains relevant for competition. In the case of utilities, the absence of a cloud–growth relationship may reflect the fact that demand is not structurally constrained. IaaS displays no significant association with growth, declining with size, in any sector, in alignment with the main results.

\subsection{Robustness checks}
\label{subsec:robustness}

\textbf{Threats to identification}. In Section \ref{sec:results}, we have estimated Equation \ref{eq:baseline_OLS} using ordinary least squares. Despite the inclusion of relevant controls and fixed effects, estimates may still be biased due to endogeneity in the relationship between cloud use and firm growth. For instance, high-growth firms may be more likely to adopt performance-enhancing technologies such as cloud computing. To address these concerns, we implement two IV specifications -- endogenous treatment and two-stage least squares, based on \citep{wooldridgeCF} -- using a measure of municipality-level lightning strike density as an instrument for cloud use. The identification strategy builds on evidence that lightning strikes increase the cost and reduce the reliability of digital infrastructure, thereby hindering broadband diffusion \citep{Gavazza2018, manacorda, Guriev2021, Goldberg2022, Caldarola2023}, a key prerequisite for cloud use \citep{Destefano2023cloud}. We measure lightning strike density at the municipality level using data from the World Wide Lightning Location Network (WWLLN) \citep{Kaplan2023}, computed as the average number of daily strikes over the period 2008-2017 and weighted by population. As lightning incidence is largely time-invariant, this measure captures persistent spatial variation in exposure rather than transitory shocks, providing plausibly exogenous variation in broadband quality and, in turn, in cloud use. Extensive details on the construction and validity of the instrument are provided in Appendix \ref{apdx_sec:identification}.

Table \ref{tab:IV} reports the results. Columns 1 and 2 present the growth and cloud equations from the endogenous treatment model. As expected, the instrument enters negatively and is highly significant, indicating that firms in areas with higher lightning density are less likely to adopt cloud technologies. The outcome equation shows that the effect of cloud use on growth is stronger for smaller firms. Columns 3-5 report TSLS estimates, which confirm the endogenous treatment results and provide statistics and test on instrument validity. The Kleibergen-Paap F-statistic exceeds 11, suggesting that the instrument is not weak, and the Anderson-Rubin test rejects the null of weak identification. Finally, in Columns 6 and 7 we experiment with a control function approach that incorporates cloud types in the growth equation and corrects for selection using the inverse Mills ratio derived from a probit model of cloud use. These estimates are in line with the baseline findings in Section \ref{sec:results}.\footnote{\rev{The TSLS coefficients on cloud use (Column 3) are considerably larger in magnitude than the corresponding OLS estimate (Table \ref{tab:OLS}, Column 4). In our case, such a gap between TSLS and OLS can arise from several sources, including a local average treatment effect concentrated among firms whose cloud use is most sensitive to broadband quality, attenuation bias in OLS from the absence of information on the exact year of first cloud use, or some degree of slack in the exclusion restriction. While the placebo tests and additional controls in Appendix \ref{apdx_sec:identification} support instrument validity, we cannot fully rule out the other possibilities, and we therefore read the TSLS magnitude with some caution.}}

We conduct an additional set of robustness checks on the causal identification of our results (Table \ref{tab:robtsls}). First, we verify that results are not driven by the ICT service sector. Second, we include year-by-region fixed effects to account for region-specific trends in the first stage. Third, we extend the growth horizon to include five-year forward window as well. Across all specifications, results remain unchanged.

\begin{table}[!htbp]\centering
    \begin{threeparttable}
        \scalebox{0.65}{
            \begin{minipage}{\textwidth}
            \centering
                \begin{tabular}{lccccccc}
                    \toprule
                    & (1) & (2) & (3) & (4) & (5) & (6) & (7) \\
                    & \multicolumn{2}{c}{Endogenous Treatment} & \multicolumn{3}{c}{Two Stage Least Square} & \multicolumn{2}{c}{IMR-augmented} \\
                    & 2nd & 1st & 2nd & 1st & 1st & 2nd & 1st \\
                    & Growth & Cloud & Growth & Cloud & Cloud $\times$ Log Sales & Growth & Cloud \\
                    \addlinespace\cmidrule(lr){2-3}\cmidrule(lr){4-6}\cmidrule(lr){7-8}\addlinespace
                    
                    Cloud (t) 
                    & 0.2885*** &  & 1.3126*** &  &  &  &  \\
                    & (0.0771)  &  & (0.4923)  &  &  &  &  \\\addlinespace
                    
                    Cloud (t)$\times$Log Sales (t-5) 
                    & -0.0170*** &  & -0.0918*** &  &  &  &  \\
                    & (0.0062)   &  & (0.0332)   &  &  &  &  \\\addlinespace
                    
                    Cloud IaaS (t)
                    &  &  &  &  &  & 0.0893 &  \\
                    &  &  &  &  &  & (0.0702) &  \\\addlinespace
                    
                    Cloud SaaS (t)
                    &  &  &  &  &  & 0.3095*** &  \\
                    &  &  &  &  &  & (0.0738)  &  \\\addlinespace
                    
                    Cloud IaaS (t)$\times$Log Sales (t-5)
                    &  &  &  &  &  & -0.0022 &  \\
                    &  &  &  &  &  & (0.0061) &  \\\addlinespace
                    
                    Cloud SaaS (t)$\times$Log Sales (t-5)
                    &  &  &  &  &  & -0.0257*** &  \\
                    &  &  &  &  &  & (0.0065)   &  \\\addlinespace
                    
                    Log Sales (t-5)
                    & -0.0421*** & 0.1546*** & -0.0290* & 0.0276** & 0.1916 & -0.0398*** & 0.1547*** \\
                    & (0.0059)   & (0.0207)  & (0.0149) & (0.0120) & (0.1287) & (0.0049)   & (0.0114)  \\\addlinespace
                    
                    Log-Lightning Strikes Density
                    &  & -0.0571*** &  & 0.0020 & 0.3731*** &  & -0.0568*** \\
                    &  & (0.0130)   &  & (0.0104) & (0.0990) &  & (0.0141)  \\\addlinespace
                    
                    Log-Lightning Strikes Density$\times$Log Sales (t-5)
                    &  &  &  & -0.0020* & -0.0603*** &  &  \\
                    &  &  &  & (0.0012) & (0.0120)   &  &  \\\addlinespace
                    
                    IMR-based Generalised Error
                    &  &  &  &  &  & 0.0009 &  \\
                    &  &  &  &  &  & (0.0170) &  \\\addlinespace
                    
                    \midrule
                    
                    Observations 
                    & 16,216 & 16,216 & 16,216 & 16,216 & 16,216 & 16,216 & 16,216 \\
                    
                    Industry, Reg. Year FE 
                    & X & X & X & X & X & X & X \\
                    
                    Firm Controls 
                    & X & X & X & X & X & X & X \\
                    
                    Municipality Controls 
                    & X & X & X & X & X & X & X \\
                    
                    KP F Statistic 
                    &  &  & 11.69 & 11.69 & 11.69 &  &  \\
                    
                    Anderson-Rubin Test 
                    &  &  & 0.0416 & 0.0416 & 0.0416 &  &  \\
                    
                    \bottomrule      
                \end{tabular}
                \begin{tablenotes}
                    \small
                    \item \textit{Note}: Columns 1-2 and Columns 3-5 correspond to an Endogenous Treatment and TSLS estimation respectively. Columns 6-7 report IMR-augmented estimates, where the selection equation is estimated via probit and the inverse Mills ratio is included in the outcome equation. Standard errors in endogenous treatment and TSLS specifications and Column 7 are clustered at the 2-digit industry level, they are based on 1000 bootstrap replications in Column 6. Firm controls are measured in $t-5$ and include the logarithms of age, tangible and intangible capital, the share of workers specialised in ICT and R\&D roles, the average hourly wage of managers and engineers, and two dummies for exporter and multi establishment status.  Standard errors are clustered clustered at NACE 2-digit level. Municipality controls include average elevation, average terrain ruggedness, a dummy for rural status, average firm growth, average firm productivity, total number of firms, total employment. The complete version of this table, including additional coefficients of controls, is reported on Table \ref{tab:IVfull} in the Appendix. *** p$<$0.01, ** p$<$0.05, * p$<$0.1.
                \end{tablenotes}
            \end{minipage}
            }
        \end{threeparttable}
        \caption{Endogenous Treatment and TSLS results}
        \label{tab:IV}
\end{table}
    
\textbf{Change in sample and variable definitions.} In Table \ref{tab:robsample}, we report four regressions aimed at assessing the robustness of our results to changes in the sample and (dependent as well as independent) variables' definitions. We first incorporate the 2020 and 2021 waves of the ICT survey, which provide information on cloud use in 2020 and 2021, in Column 1. We next exclude ICT service firms in Column 2 -- a common practice in studies of digital technologies, as firms in these sectors are typically more directly involved in the production and provision of digital services and may therefore differ in their use of such technologies \citep{Calvino2026respol}. Column 3 uses employment as an alternative measure of firm size, both to compute growth rates and to measure initial size. Employment is less sensitive to demand fluctuations than sales and may capture a distinct dimension of firm size, particularly in the context of digital technologies that may affect labour and output differently \citep[see e.g.][]{bisiocuzzola2023}. Finally, in Column 4, we disaggregate IaaS cloud services into two categories: computing and storage. Across all specifications, the results remain consistent with the baseline findings.

\textbf{Quantiles and bins of size and age.} We explore potential non-linearities in firm size and age. First, Table \ref{tab:robquantiles} reports estimates of Equation \ref{eq:baseline_OLS} when firms are divided into sector-year-specific quartiles of sales and age. In Columns 1 and 2, we sequentially replace the continuous measures with sales quartiles and then age quartiles. The results indicate that the association between SaaS cloud use and firm outcomes varies systematically across size quartiles, while the interaction with age is only statistically significant in the highest age quartile.
Second, Table \ref{tab:robclass} reports results based on alternative discrete classifications of firm size and age. Column 1 uses the standard Eurostat size classification\footnote{Small: less than 50 employees; Medium: between 50 and 250 employees; Large: more than 250 employees.} and a commonly used age classification,\footnote{Startup: less than 5 years; Young: between 6 and 10 years old; Mature: more than 11 years old.} while Column 2 further refines these categories by distinguishing additional size classes\footnote{Between 250 and 500; between 500 and 1000 employees; and more than 1000 employees.} and age classes.\footnote{ Between 11 and 25 years old; between 26 and 50 years old; and more than 50 years old.} The results confirm the baseline findings; the interaction between SaaS cloud use and firm size remains negative and statistically significant across size classes, while the interactions with age classes are generally not significant.

\textbf{The sales growth period.} This robustness check involves varying the time window around time $t$ (the time of first reported cloud use), departing from the specification defined in Equations \ref{eq:baseline_OLS}. This has implications for the distance between time $t$ and the start and end points of the growth period. Results are reported in Tables \ref{tab:robforwardgrowth}, \rev{\ref{tab:forward2}} and \ref{tab:roblength} in Appendix \ref{apdx_sec:full_tables}. In Table \ref{tab:robforwardgrowth}, we extend the forward-looking growth window. In particular, we compute growth rates starting at $t-5$, in order to capture potential early effects of cloud use, and ending at alternative horizons ranging from $t+1$ to $t+5$, to account for firms whose performance effects may materialise with a delay, e.g. late adopters. The results remain broadly consistent with those in Table \ref{tab:OLS}, where the growth period spans from $t-5$ to $t$. \rev{Next, in Table \ref{tab:forward2}, we restrict the growth window to forward-looking pre-pandemic horizons, retaining the longest window available per ICT survey wave: $t$ to $t+3$ for 2016 (through 2016--2019), and $t$ to $t+1$ for 2018 (through 2018--2019). The 2016--2019 estimates confirm our main results. The 2018--2019 estimates are not significant, consistent with a window too short to capture effects on growth. However, shifting to 2017--2019 restores significance.\footnote{\rev{The measurement error is limited, as cloud use is measured at early 2018, given the survey's early-year timing. See Section \ref{sec:data}.}} Finally}, in Table \ref{tab:roblength}, we vary the length of the growth window by shrinking (Column 1) and extending (Column 2) the baseline period by two years. Once again, the results remain consistent with baseline findings.

\textbf{Long-differences equations.} Next, we adopt a long-differences approach, which is frequently used in studies examining ICT impacts \citep[e.g.,][]{acemoglurestreporobot,babina2021,forman2012}. We report results in Table \ref{tab:roblongdiff}. This method enables us to estimate the effects of cloud use over its entire diffusion period, assumed to have begun in 2009 following \citet{bloomcloud2018}, despite taking place within each firm at different points in time. This approach implies the use of one ICT survey at a time. We implement the following cross-sectional regression:
\begin{equation}\label{eq:longdiff}
\begin{split}
  & \text{Sales Growth}_{i, t, 2009} = \\
  & a + \beta_1 \Delta \text{Cloud}_{i}  +\beta_2 \Delta \text{Cloud}_{i,t} \cdot \text{Log-Sales}_{i,2009} + \beta_3\text{Log-Sales}_{i,2009}+ \beta_X\text{X}_{i,2009} + \epsilon_{i}
\end{split} 
\end{equation}
\noindent Here, $\Delta \text{Cloud}_{i}$ is a dummy variable for cloud use in year $t$, indicating whether cloud technology was adopted between 2009 and $t$ (2016, or 2018). Since cloud use was unlikely in 2009, $\Delta \text{Cloud}_{i}$ effectively captures the transition to cloud use by firms \citep[similarly to the identification strategy in][]{formanKM, forman2012}. Table \ref{tab:roblongdiff} shows that the results for the long run growth between 2009 and 2018 (Column 1) or between 2009 and 2016 (Column 2) are consistent with the main findings.

\section{Mechanisms}
\label{sec:mechanism}

The main results presented in the previous section have demonstrated the positive relationship between cloud use and firms' growth rates, with smaller firms benefiting to a larger extent. We have also shown that this effect is mainly driven by SaaS cloud.

\rev{\textbf{ICT capabilities complementarity.} A central theoretical ambiguity in the cloud-growth relationship concerns whether SaaS use substitutes also for other firms' ICT capabilities in the firm -- with firms entirely outsourcing their ICT capacity to cloud providers -- or if instead it complements them. We test these competing mechanisms by estimating Equation~\ref{eq:baseline_OLS} on subsamples defined by the presence of ICT capabilities, measured by the use of at least one other digital technology in the ICT survey, and reported in Table~\ref{tab:SaaSICT}. Columns 1-4 and 5-8 report split-sample estimates for firms with and without ICT capabilities, respectively, with controls, industry, and region fixed effects added incrementally. The positive association between SaaS cloud and growth, decreasing with firm size, is significant across specifications only in the subsample of ICT-capable firms (Columns 1-4). Among firms without ICT capabilities, the SaaS coefficients are smaller in magnitude, and its interaction with size is not statistically significant at any specification (Columns 5-8).}

\rev{To formally assess whether this size gradient differs across the two groups, Column 9 reports a pooled specification with the full set of triple interactions between cloud type, ICT capability, and log sales. The triple interaction between SaaS cloud, ICT capability, and firm size is negative and significant, while the corresponding IaaS interaction is not. Similarly, the coefficient of the cloud-ICT capabilities interaction is positive and significant only for SaaS. This indicates that the size-contingent SaaS-growth association is significantly stronger among ICT-capable firms. This is corroborated by a Wald test comparing the SaaS $\times$ Log Sales coefficients across the fully specified split-sample models (Columns 4 and 8), which rejects equality of the two coefficients (p=0.031).}

\rev{This result resolves the ambiguity in favour of the ICT complementarity interpretation. SaaS widens access to digital capabilities, but the disproportionate returns that smaller firms derive from that access are conditioned on ICT complementary capabilities, as proxied by the use of other ICT technologies. The implication is that SaaS shifts the binding constraint on firm growth from technology access to digital readiness (measured by the stock of ICT capabilities and complementary assets) without eliminating the underlying heterogeneity in firms' capacity to exploit digital inputs.}\footnote{\rev{We also considered the presence of ICT workers as an alternative measure. Even though results mirrored ones in Columns 4 and 8 of Table~\ref{tab:SaaSICT}, with the interaction term being significant only for ICT-capable firm, the pooled triple-interaction specification based on this measure did not yield a significant cloud SaaS $\times$ Log Sales $\times$ ICT-workers coefficient. This may depend on the nature of these two proxies of digital capabilities. The digital-technology measure is conceptually closer to the notion of ICT capability that the mechanism requires, that is the use of other digital technologies that cloud provide or improve access to, while as the ICT-worker indicator captures a narrower, human-capital-specific dimension of ICT capabilities, possibly not considering the use of other IT tools.}}

\begin{table}[!htbp]\centering
    \begin{threeparttable}
        \scalebox{0.5}{\color{black}
            \begin{minipage}{\textwidth}
                \centering
                \begin{tabular}{lccccccccc}
                    \toprule
                    & (1)        & (2)        & (3)        & (4)        & (5)        & (6)        & (7)        & (8)        & (9)          \\
                    & \multicolumn{4}{c}{Firms with ICT capabilities} & \multicolumn{4}{c}{Firms without ICT capabilities} & Triple Interaction \\\addlinespace\cmidrule(lr){2-5}\cmidrule(lr){6-9}\cmidrule(lr){10-10}\addlinespace
                    IaaS Cloud (t)                                            & 0.0673     & 0.0686     & 0.0185     & 0.0209     & 0.2163**   & 0.2054**   & 0.1688**   & 0.1304     & 0.1316       \\
                    & (0.1158)   & (0.1167)   & (0.1228)   & (0.1164)   & (0.0871)   & (0.0855)   & (0.0771)   & (0.0802)   & (0.0822)     \\\addlinespace
                    SaaS Cloud (t)                                            & 0.5710***  & 0.5665***  & 0.5127***  & 0.4923***  & 0.2042*    & 0.2109**   & 0.1366     & 0.1288     & 0.1278       \\
                    & (0.1407)   & (0.1383)   & (0.1382)   & (0.1302)   & (0.1072)   & (0.1053)   & (0.0957)   & (0.0989)   & (0.0959)     \\\addlinespace
                    Log Sales (t-5)                                           & -0.0352*** & -0.0354*** & -0.0457*** & -0.0331*** & -0.0353*** & -0.0364*** & -0.0449*** & -0.0465*** & -0.0424***   \\
                    & (0.0065)   & (0.0063)   & (0.0052)   & (0.0100)   & (0.0050)   & (0.0053)   & (0.0049)   & (0.0061)   & (0.0061)     \\\addlinespace
                    IaaS Cloud (t)$\times$Log Sales (t-5)                     & 0.0010     & 0.0005     & 0.0046     & 0.0033     & -0.0131    & -0.0121    & -0.0097    & -0.0069    & -0.0069      \\
                    & (0.0100)   & (0.0100)   & (0.0105)   & (0.0100)   & (0.0084)   & (0.0082)   & (0.0079)   & (0.0082)   & (0.0084)     \\\addlinespace
                    SaaS Cloud (t)$\times$Log Sales (t-5)                     & -0.0479*** & -0.0475*** & -0.0430*** & -0.0416*** & -0.0152    & -0.0162    & -0.0099    & -0.0096    & -0.0094      \\
                    & (0.0111)   & (0.0109)   & (0.0108)   & (0.0104)   & (0.0099)   & (0.0097)   & (0.0091)   & (0.0095)   & (0.0093)     \\\addlinespace
                    ICT Capabilities (t)                                      &            &            &            &            &            &            &            &            & 0.0312       \\
                    &            &            &            &            &            &            &            &            & (0.0539)     \\\addlinespace
                    IaaS Cloud (t)$\times$ICT Capabilities (t)                &            &            &            &            &            &            &            &            & -0.0956      \\
                    &            &            &            &            &            &            &            &            & (0.1482)     \\\addlinespace
                    SaaS Cloud (t)$\times$ICT Capabilities (t)                &            &            &            &            &            &            &            &            & 0.3518**     \\
                    &            &            &            &            &            &            &            &            & (0.1731)     \\\addlinespace
                    ICT Capabilities (t)$\times$Log Sales (t-5)               &            &            &            &            &            &            &            &            & 0.0030       \\
                    &            &            &            &            &            &            &            &            & (0.0053)     \\\addlinespace
                    IaaS Cloud (t)$\times$ICT Capabilities (t)$\times$Log Sales (t-5) &     &            &            &            &            &            &            &            & 0.0091       \\
                    &            &            &            &            &            &            &            &            & (0.0137)     \\\addlinespace
                    SaaS Cloud (t)$\times$ICT Capabilities (t)$\times$Log Sales (t-5) &     &            &            &            &            &            &            &            & -0.0312**    \\
                    &            &            &            &            &            &            &            &            & (0.0150)     \\
                    \addlinespace\midrule\addlinespace
                    Observations                                               & 6,592      & 6,592      & 6,592      & 6,592      & 9,624      & 9,624      & 9,624      & 9,624      & 16,216       \\
                    Year FE                                                    & X          & X          & X          & X          & X          & X          & X          & X          & X            \\
                    Region FE                                                  &            & X          & X          & X          &            & X          & X          & X          & X            \\
                    Industry FE                                                &            &            & X          & X          &            &            & X          & X          & X            \\
                    Other Controls                                             &            &            &            & X          &            &            &            & X          & X            \\
                    Adj R2                                                     & 0.0630     & 0.0652     & 0.0934     & 0.127      & 0.0335     & 0.0363     & 0.0659     & 0.0911     & 0.103        \\\bottomrule
                \end{tabular}
                \begin{tablenotes}
                    \footnotesize
                    \item \textit{Note}: Split sample estimations of Equation \ref{eq:baseline_OLS}. Columns 1-4 and 5-8 split the sample based on the presence of ICT capabilities, as measured by the use of other digital technologies in $t$. Column 9 includes the saturated interactions between cloud types, ICT capabilities and size. Controls are measured in $t-5$ and include the logarithms of age, tangible and intangible capital, the share of workers specialised in ICT and R\&D roles, the average hourly wage of managers and engineers, and two dummies for exporter and multi establishment status. Standard errors are clustered at NACE 2-digit level. The complete version of this table, including additional coefficients of controls, is available upon request. *** p$<$0.01, ** p$<$0.05, * p$<$0.1.
                \end{tablenotes}
            \end{minipage}}
    \end{threeparttable}
    \caption{SaaS cloud complementarity with ICT capabilities}
    \label{tab:SaaSICT}
\end{table}

\textbf{Cloud SaaS organisational changes.} Having established that SaaS cloud drives growth primarily among ICT-capable firms, we now examine whether and how its use is accompanied by internal reorganisation of workforce composition. We test the hypotheses that SaaS cloud use is associated with occupational upgrading and with an expansion of upper-layer workforce shares using changes in workforce composition, measured by the share of workers across occupational categories as dependent variables in Equation \ref{eq:baseline_OLS}. We distinguish four categories: ICT technical occupations (ICT engineers and technicians), non-ICT technical occupations (non-ICT engineers and technicians), non-technical executives and intermediate occupations, and manual and clerical workers.\footnote{ICT technical occupations include occupations classified under codes 388a-388e and 4781-478d of the PCS 2003 classification. Non-ICT technical occupations include all codes in Sections 38 (engineers) and 47 (technicians), excluding ICT-related codes. The executives and intermediates category includes Sections 3 and 4 of the same classification, net of the codes listed above. The clerical and manual workers category includes Sections 5 and 6 of the same classification.} 

Results are reported in Table~\ref{tab:SaaSwork}. SaaS cloud is associated with a reallocation of workers toward higher-skill layers. Both the share of ICT occupations (Column~1) and non-technical executives and intermediate workers (Column~3) are positively related to SaaS cloud use, with this association weakening for larger firms. This suggests that the barriers to adding upper-layer workers \citep[documented in growing firms,][]{caliendo2015organization} may bind more tightly at smaller scales, making SaaS-enabled reorganisation particularly relevant there. The absence of an association with non-ICT technical workers (Column~2) suggests that SaaS reorganisation is not a generic upskilling, but a reallocation specifically toward digital and managerial/intermediate functions. These shifts are mirrored by a crowding out of manual and clerical occupations (Column~4).

These results suggest that SaaS cloud use does not merely improve performance, but reshapes the internal organisation of firms. This pattern is consistent with SaaS enabling the organisational changes necessary to digitalise and scale, particularly among smaller firms where such reorganisation may be more substantial.  

\begin{table}[!htbp]\centering
    \begin{threeparttable}
        \scalebox{0.65}{
            \begin{minipage}{\textwidth}
                \centering
                \begin{tabular}{lcccc}
                    \toprule
                    & (1)                                         & (2)                 & (3)     & (4)                      \\
                    & ICT Technical & Non-ICT Technical &  Executives/intermediates &  Manual/Clerical \\
                    \addlinespace\cmidrule(lr){2-5}\addlinespace
                    IaaS Cloud (t)                        & -0.0028               & 0.0071                    & 0.0219              & -0.0246                  \\
                    & (0.0106)              & (0.0150)                  & (0.0175)            & (0.0151)                 \\\addlinespace
                    SaaS Cloud (t)          & 0.0213*               & 0.0134                    & 0.0432**            & -0.0707***               \\
                    & (0.0124)              & (0.0155)                  & (0.0182)            & (0.0170)                 \\\addlinespace
                    IaaS Cloud (t)$\times$Log Sales (t-5) & 0.0004                & -0.0003                   & -0.0025             & 0.0022                   \\
                    & (0.0010)              & (0.0014)                  & (0.0016)            & (0.0014)                 \\\addlinespace
                    SaaS Cloud (t)$\times$Log Sales (t-5) & -0.0017*              & -0.0010                   & -0.0039**           & 0.0061***                \\
                    & (0.0010)              & (0.0014)                  & (0.0017)            & (0.0016)                 \\\addlinespace
                    Log Sales (t-5)                       & 0.0002                & 0.0030*                   & 0.0099***           & -0.0126***               \\
                    & (0.0010)              & (0.0016)                  & (0.0033)            & (0.0025)                 \\
                    \addlinespace\midrule\addlinespace
                    Observations                          & 16,216                & 16,216                    & 16,216              & 16,216                   \\
                    Industry, Reg., Year FE               & X                     & X                         & X                   & X                        \\
                    Controls                              & X                     & X                         & X                   & X                        \\
                    Adj R2                                & 0.0907                & 0.0595                    & 0.0325              & 0.0482                   \\\bottomrule
                \end{tabular}
                \begin{tablenotes}
                    \footnotesize
                    \item \textit{Note}: Dependent variable in Columns 1-4 is the change in the share of hours worked by ICT engineers, executives, Non-ICT engineers and manual or clerical workers, respectively. Controls are measured in $t-5$ and include the logarithms of age, tangible and intangible capital, the share of workers specialised in ICT and R\&D roles, the average hourly wage of managers and engineers, and two dummies for exporter and multi establishment status. Standard errors are clustered at NACE 2-digit level. The complete version of this table, including additional coefficients of controls, is available upon request. *** p$<$0.01, ** p$<$0.05, * p$<$0.1.
                \end{tablenotes}
            \end{minipage}
        }
    \end{threeparttable}
    \caption{SaaS cloud and workforce reorganisation}
    \label{tab:SaaSwork}
\end{table}

\section{Cloud and industry concentration}
\label{sec:concentration}

In Section \ref{sec:results} we have demonstrated that the use of cloud services positively impacts firms' long-term sales growth rate, with smaller firms growing more by leveraging cloud technologies to narrow their digital gap. This finding talks directly to a growing literature examining the link between ICT and industry concentration \citep{bajgar2021,brynjolfssonICTconc2023, babina2021, deridderaer}. It is therefore an open question whether the widespread use of cloud technologies may be associated with a moderation of the increasing concentration trends observed across high-income markets \citep{bajgar2019}.

We provide suggestive and descriptive evidence in this direction, consistent with the firm-level results presented in the previous section. To do so, we aggregate our firm-level data by 2-digit industrial sectors and estimate the following Equation \citep{bessenindconc2020,brynjolfssonICTconc2023,mcafee2008investing}:

\begin{equation}\label{eq:concentration}
\begin{split}
  & \text{Log Concentration}_{s,t} = \\
  & \alpha + \beta_1 \text{Cloud Share}_{s,t} + \beta_x \text{Controls}_{s,t}  + \text{2-digit Ind.}_{s} + \text{Year}_{t} + \epsilon_{s,t}
\end{split}
\end{equation}

Where $s$ identifies 2-digit sectors at time $t$ (either 2016 or 2018), $\text{Concentration}_{s,t}$ is the Herfindahl-Hirschman Index (HHI) of 2-digit industries at each time period, $\text{Cloud Share}_{s,t}$ is the sectoral share of firms using cloud at time $t$, $\text{Controls}_{i,t}$ is a vector including the the stock ICT and R\&D Full Time Equivalent (FTE) occupations, the share of firms using big data analytics and participating in e-commerce activities, the aggregate sales and employment of active firms, aggregate physical and intangible capital stocks, and the number of firms in sector $s$ in logarithmic scale. We include also industry ($\text{2-digit Ind.}_{s}$) and year ($\text{Year}_{t}$) fixed effects.\footnote{We observe industries at the 2-digit level, as the limited number of firms in the sectoral samples provided by the ICT surveys does not allow us to measure the share of firms purchasing cloud services at lower levels of aggregation. Moreover, due to the small number of observations included in this sample, results should be interpreted with caution.}

\begin{table}[!htbp] \centering 
    \begin{threeparttable}
        \scalebox{0.65}{
            \begin{minipage}{\textwidth}
                \begin{tabular}{lccccccc}
                    \toprule
                    & (1)           & (2)      & (3)        & (4)       & (5)       & (6)        & (7)       \\
                    \addlinespace\cmidrule(lr){2-8}
                    Cloud Share                    & 0.1147        & -0.0627  & -0.2460**  & -0.3859** & -0.4472** & -0.4853*** & -0.5281*  \\
                    & (0.0808)      & (0.2126) & (0.1094)   & (0.1748)  & (0.1793)  & (0.1701)   & (0.2875)  \\\addlinespace
                    E-commerce Share               &               & -0.2850  &            & -0.2236   & -0.1908   & -0.1221    & -0.1649   \\
                    &               & (0.2637) &            & (0.2426)  & (0.2698)  & (0.2703)   & (0.4093)  \\\addlinespace
                    Big Data Share        &               & 0.1155   &            & 0.1274    & 0.0771    & 0.1428     & 0.3594    \\
                    &               & (0.2088) &            & (0.2159)  & (0.2343)  & (0.2126)   & (0.2884)  \\\addlinespace
                    Log ICT FTE   Occupations & & 0.2704   &            & 0.0160    & 0.0200    & 0.0868     & -0.0100   \\
                    &               & (0.3940) &            & (0.1678)  & (0.1618)  & (0.1731)   & (0.1764)  \\\addlinespace
                    Log Total Employment           &               &          &            &           & -0.9975   & -0.8781    & -1.5766   \\
                    &               &          &            &           & (0.6885)  & (0.7907)   & (0.9924)  \\\addlinespace
                    Log Total Sales                &               &          & 1.5843***  & 1.5960*** & 2.0448*** & 2.0246***  & 2.2972*** \\
                    &               &          & (0.1323)   & (0.1476)  & (0.3369)  & (0.3395)   & (0.4965)  \\\addlinespace
                    Log Number of Firms            &               &          & -0.6863*** & -0.6464** & -0.4420   & -0.6063*   & -0.6158   \\
                    &               &          & (0.2487)   & (0.2745)  & (0.3009)  & (0.3537)   & (0.3987)  \\\addlinespace
                    Log R\&D   FTE Occupations& & -0.2542  &            & -0.1321   & -0.1702   & -0.2140*   & -0.0879   \\
                    &               & (0.2863) &            & (0.1237)  & (0.1089)  & (0.1100)   & (0.1438)  \\\addlinespace
                    Log Intagible Capital          &               &          &            &           & 0.1386    & 0.0257     & 0.5730    \\
                    &               &          &            &           & (0.4349)  & (0.5200)   & (0.4728)  \\\addlinespace
                    Log Physical Capital           &               &          &            &           & -0.1681   & -0.0869    & -0.2891   \\
                    &               &          &            &           & (0.3105)  & (0.3324)   & (0.4257)  \\
                    \addlinespace\midrule\addlinespace
                    Observations                   & 120           & 120      & 120        & 120       & 120       & 108        & 120       \\
                    Sector-Year FE                 & X           & X      & X        & X       & X       & X        & X       \\
                    Adj. R2                         & 0.970         & 0.969    & 0.988      & 0.987     & 0.986     & 0.986      & 0.699    \\\bottomrule
                \end{tabular}
                \begin{tablenotes}
                    \item \textit{Note}: The table reports the results of the estimation of Equation \ref{eq:concentration}. The dependent variable measures the Herfindahl-Hirschman Index (HHI) of 2-digit NACE industries, obtained by aggregating the sales of firms by industry in Columns 1-6, and its 5-years backward growth rate in Column 7. The cloud, e-commerce and big data shares measure the share of firms buying cloud services, selling goods via e-commerce or using big data analytics. Controls are measured at time $t$ and include: the log of the number of employees in ICT and R\&D occupations; the share of firms using BDA and participating in e-commerce activities; the log of total revenues of the industry (in Euros); the log number of total employees and number of firms; the log of the total stock of tangible and intangible capital (in Euros). Column 6 excludes ICT service sectors. All regressions are estimated using OLS. Standard errors are clustered at the NACE 2-digit level. *** p$<$0.01, ** p$<$0.05, * p$<$0.1.
                \end{tablenotes}
            \end{minipage}
        }
    \end{threeparttable}
    \caption{Cloud and sectoral concentration}
    \label{tab:concentration}
\end{table}

Table \ref{tab:concentration} presents the estimation results of Equation \ref{eq:concentration}. In the baseline specification without controls (Column 1), we observe a positive relationship between industry concentration and the share of firms adopting cloud services. However, this relationship turns negative once ICT and R\&D controls (Column 2) or sectoral size controls (Column 3) are included. This pattern suggests that the initial positive association may be driven by the correlation between cloud use and ICT intensity, the latter of which have been shown to be positively linked to concentration \citep{bessenindconc2020, brynjolfssonICTconc2023}. As expected, the number of firms and total sales in a sector are negatively and positively associated with concentration, respectively. This supports the idea that the positive association in Column 1 may be confounded by sectors experiencing rapid growth in both firm count and sales -- such as digital-intensive industries \citep{bajgar2021}, which may also be undergoing faster digitalisation. Finally, in Column 4, we include both sets of controls from Columns 2 and 3, and in Column 5 we further control for capital and employment. In both cases, the relationship between cloud use and industry concentration remains negative and statistically significant. 
Column 6 reports estimates of Equation \ref{eq:concentration} excluding the ICT services sector from the sample; the negative association between cloud use and concentration is preserved, although its magnitude should be interpreted in light of the reduced sample. Finally, in Column 7 we use the 5-year growth rate in concentration as the dependent variable. 
While this pattern is robust to the inclusion of additional sectoral characteristics, to changes in sample and in dependent variable, it should be interpreted as an association, and we remain cautious about attributing a causal interpretation.

Overall, the observed negative relationship between the share of cloud users in an industry and its concentration is consistent with the idea that cloud services may be associated with relatively stronger growth among smaller firms compared to larger ones, and that this pattern at the firm level could be linked to a modest reduction in industrial concentration.

\section{Conclusions}
\label{sec:conclusion}

\rev{This paper documents an exception to the prevailing pattern of ICT-driven growth. Smaller firms derive larger growth benefits from cloud use than their larger counterparts -- an effect driven entirely by SaaS cloud, whereas cloud services used for storage or computing power show no comparable size-contingent effect. However, our findings support a \textquote{conditional cloud exception} narrative, whereby the growth advantage from cloud use accrues to smaller firms when they possess complementary ICT capabilities. This suggests that while SaaS cloud broadens access to digital technologies, its effective use is conditional on firms' ICT readiness.} In addition, we show that SaaS cloud use by smaller firms is associated with an internal reorganisation of the workforce toward ICT workers and executive and intermediate occupations, consistent with the view that SaaS cloud operates as an organisational innovation enabling small firms to digitalise and scale up. Finally, we present descriptive evidence showing that, in line with the micro-level findings, industries where cloud use is more frequent tend to be less concentrated. 

These findings carry important policy implications. First, the evidence points to substantial untapped growth potential among smaller firms \rev{that possess or develop complementary ICT capabilities}. That could be unlocked through broader diffusion of SaaS cloud technologies. This is particularly relevant in Europe, where firm size distributions are skewed toward small firms and scaling frictions are pronounced \citep{garicanolelargereenen2016}, while existing evidence on age-related effects largely relies on non-European data \citep{Destefano2023cloud,wangjincloud2024}. Second, \rev{precisely because effective use depends on these capabilities,} our results suggest that effective diffusion policies must extend beyond facilitating technology acquisition alone \citep{Calvino2026respol} providing access to the additional ICT capabilities necessary to leverage SaaS cloud technologies. \rev{Put differently, our findings suggest that diffusion and capability-building policies are complements, not substitutes. Subsidising cloud diffusion alone is unlikely to activate the growth margin we document among firms lacking complementary capabilities, and may even widen digital divides within the small-firm population.} Third, these results speak directly to the well-documented rise in industry concentration across many countries \citep{gutierrez2017, bajgar2019, grullon2019, autor2020}, where ICT use and the diffusion of intangible assets have been identified as contributing \citep{bessenindconc2020, bajgar2021, brynjolfssonICTconc2023, Lashkari2024, chiavari}, in line with the idea that firms may not have equal access to new technologies \citep{terranova, delligatti2023mind}. To the extent that SaaS cloud disproportionately benefits smaller firms \rev{with ICT capabilities}, its wider diffusion \rev{paired with policies that build those capabilities} may help counteract these concentrating forces while simultaneously advancing the digital transformation of the broader firm population.

More broadly, understanding how new digital technologies interact with firm-specific characteristics remains an important avenue for future research. Future work could investigate the implications of the Industry 4.0 transition beyond cloud services, examining whether the patterns documented here generalise to other emerging digital tools. Technologies such as AI and 3D printing have the potential to support the generation and validation of new ideas \citep{nambisanrespol, beltagui, cockburn2018impact}, functioning as an \textquote{invention of a method of invention} \citep{griliches1957}. On the one hand, by lowering the barriers to new firm creation, such technologies may reshape entrepreneurship and contribute to the understanding of firm-level drivers of business dynamism, which has been linked to digital intensity at the sectoral level \citep{calvino2019dyn, calvino2020dyn}. On the other hand, R\&D capabilities and proprietary software are key determinants of performance among large firms \citep{bessenindconc2020, babina2021, coadquantile}, suggesting that more complex digital technologies may yield heterogeneous returns depending on firm characteristics and potentially reinforcing productivity divergence \citep{corrado2021} beyond concentration and business dynamism trends, rather than mitigating it.

\newpage
\small

\clearpage
\phantomsection
\addcontentsline{toc}{chapter}{Bibliography}

\footnotesize
\singlespacing

\bibliography{cloud_paper_refs.bib} 

\newpage

\appendix 

\numberwithin{equation}{section}
\numberwithin{table}{section}
\numberwithin{figure}{section}


\section{Details about identification strategy}
\label{apdx_sec:identification}

\setcounter{figure}{0}
\renewcommand{\thefigure}{A\arabic{figure}}

\setcounter{table}{0}
\renewcommand{\thetable}{A\arabic{table}}
\renewcommand{\theHtable}{table.A\arabic{table}} 

\normalsize
\onehalfspacing

To mitigate the endogeneity concerns around the relationship between cloud use and firms' growth we estimate endogenous treatment and Two-Stages Least Squares (TSLS) models that exploit a source of spatial exogenous variation associated to the cost and quality of ICT investments. We build on evidence on the US by \citet{Andersen2012} showing that ICT investments are slowed down by the incidence of a natural phenomenon: lightning strikes.\footnote{The argument proposed by Andersen is based on the correlation between lightning strike density and productivity across US states that emerged in the 1990s, when ICTs started to diffuse more widely.} By causing energy spikes and dips, lightning strikes increase the cost of digital infrastructures and technologies, slowing down their diffusion. In particular, it has been shown that the incidence of lightning strikes reduces the quality of broadband connections \citep{Yu2023}, with broadband network failures being four times more likely during thunderstorms \citep{Schulman2011}. However, in order to adopt cloud technologies, firms need to have access to reliable, fast, and state-of-art internet connection \citep{nicoletti2020, garrison, ohnemus}, which is enabled by the presence of physical infrastructure providing high-quality broadband network. Precisely, \citet{Destefano2023cloud} identify cloud use by exploiting exogenous variation in access to broadband internet. Their instrument is based on a dummy variable measuring the enabled access to fibre broadband at a given time, and the distance between firms and the closest telephone exchange site. Absent this information in our data, we resort to an instrument that exogenously predicts the diffusion of broadband internet, which in turn is more likely to be available in areas with lower density of lightning density \citep[see][]{Gavazza2018, manacorda, Guriev2021, Goldberg2022, Caldarola2023}. 

\begin{figure}[!ht]
    \centering
    \includegraphics[width=1\linewidth]{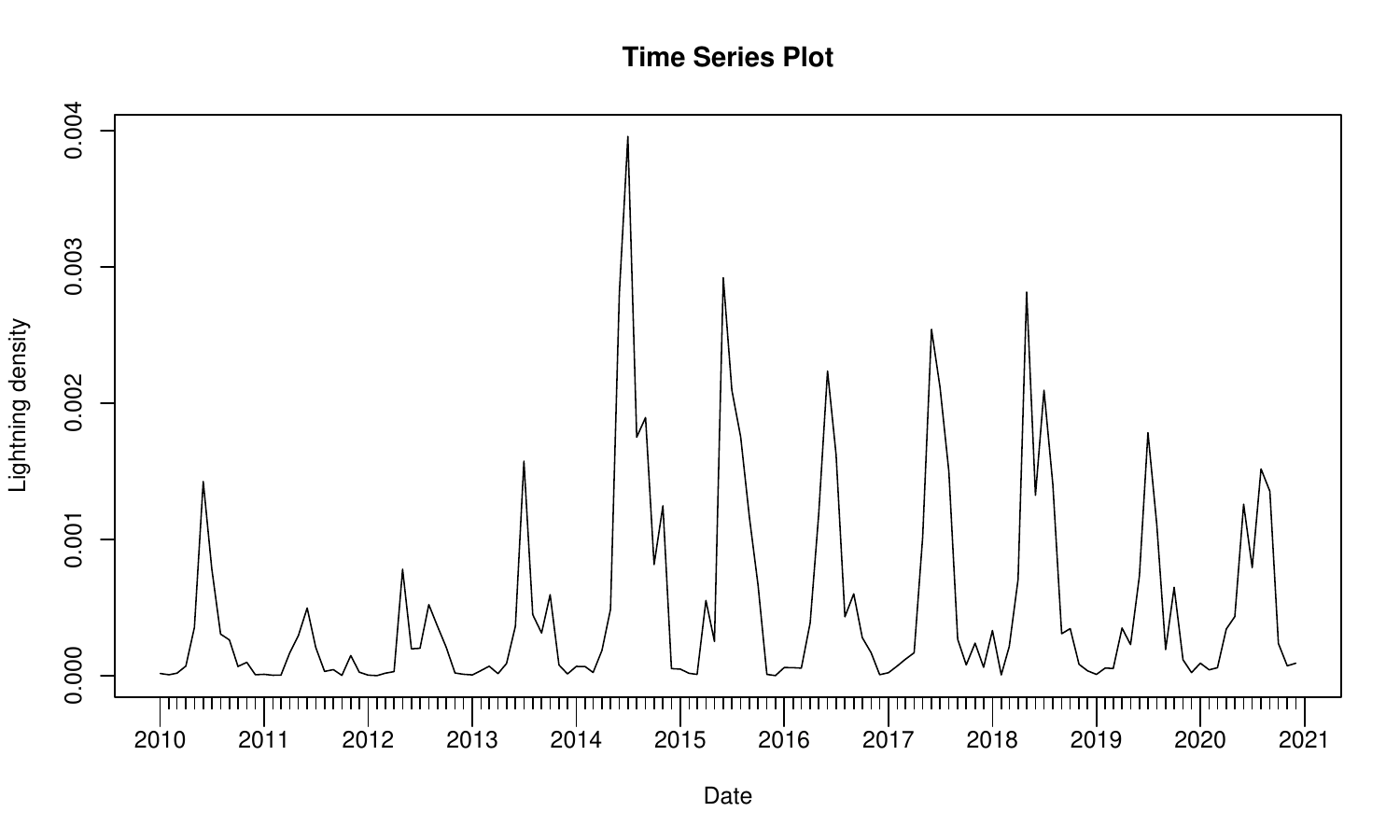}
    \caption{Lightning density monthly time series for France, 2010-2021}
    \label{fig:timeseries}
\end{figure}

Information on lightning strike density is obtained from the World Wide Lightning Location Network (WWLLN) Global Lightning Climatology and Time Series \citep{Kaplan2023}. The raw WWLLN data is a grid of 5-arcminute cells, each containing information on the count of daily lightning strikes per square kilometre. Data collection occurs daily and spans from 2008 to 2020. Building on evidence that the incidence of lightning strikes is a stationary phenomenon (see Table \ref{tab:stationarity_tests}, Figure \ref{fig:timeseries}, and \citealt{Andersen2012}), we are interested in constructing a measure that captures the geographical exposure of geographical areas to this phenomenon (in our case, the French municipality where firms are located).\footnote{French municipalities, or communes, are the smallest administrative subdivision in France, acting as local authorities. There are 34,826 communes in the country.} We calculate the average lightning strike density for each French municipality, based on the density values contained in the cells that fall within each municipality over the period 2008-2017,\footnote{In the case of cells that fall over a border between two or more communes, the cropped cells weight in each commune based on the percentage of the cell falling within each of them.} ending one year before the last survey round in our data. The resulting quantity measures the average daily lightning strikes per square kilometre in each municipality. This is then multiplied by the surface area to retrieve the total amount of lightning strikes, and weighted by population.\footnote{We source the population data at the municipality level from French official statistical sources at this link: \url{https://www.insee.fr/fr/statistiques/3698339}.}

\begin{figure}
    \centering
    \includegraphics[width=0.8\linewidth]{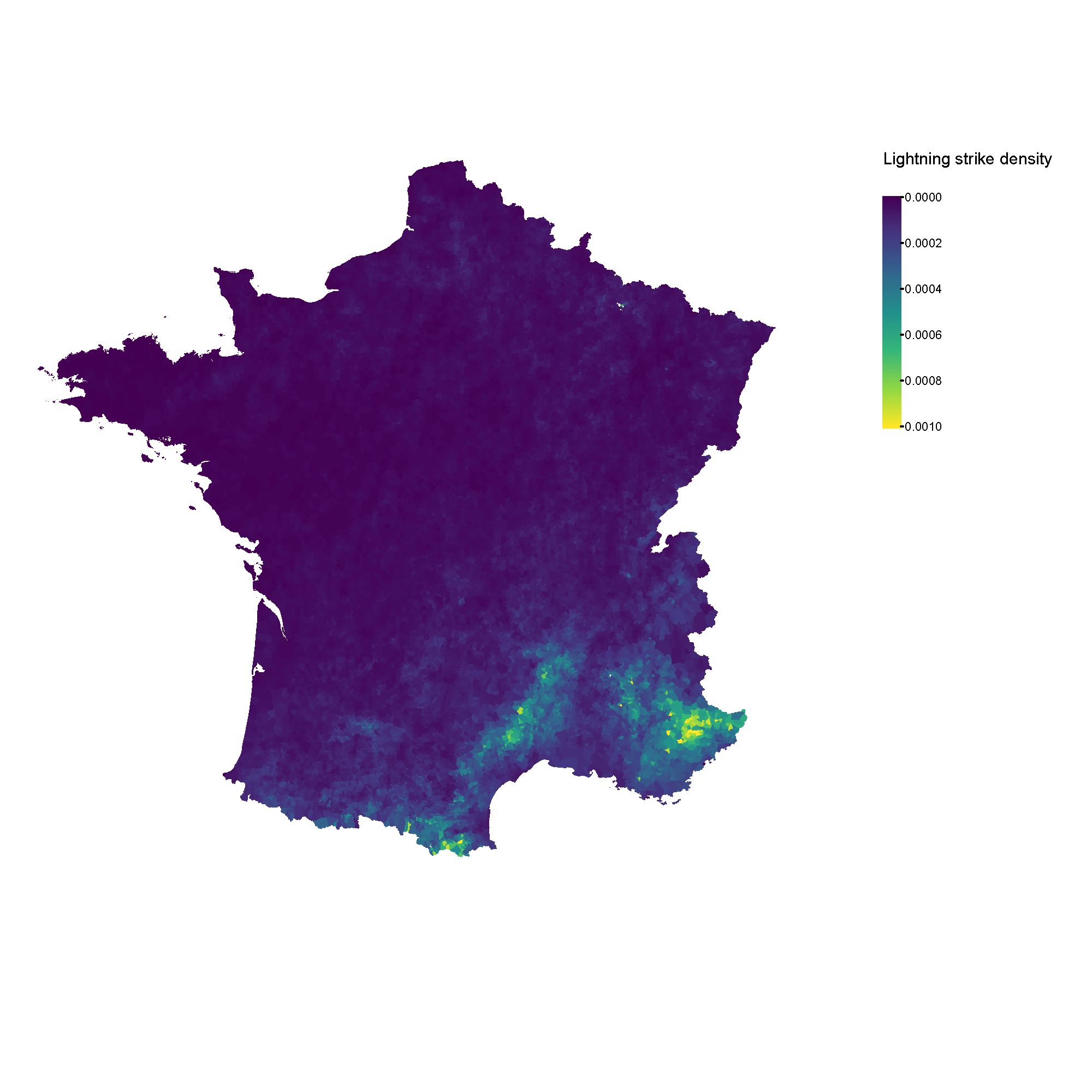}
    \caption{The maps shows the geographical distribution of the instrumental variable, which measures average daily lightning strikes at municipality level, weighted by population.}
    \label{fig:lightningpc_figure}
\end{figure}

We report the geographical distribution of our instrument in Figure \ref{fig:lightningpc_figure}.
The instrument displays substantial heterogeneity across municipalities, with a clear north-south divide. Municipalities located in the regions of Auvergne-Rhône-Alpes, Provence-Alpes-Côte d’Azur, Occitanie, and the southern part of Nouvelle-Aquitaine experience a higher density of lightning strikes. Notably, several major French technological and industrial clusters are situated within these areas, including the aerospace hub around Toulouse (Occitanie), the microelectronics and digital manufacturing industries in the Lyon-Grenoble area (Auvergne-Rhône-Alpes), the ICT poles of Aix-Marseille and Sophia Antipolis (Provence-Alpes-Côte d’Azur), and the aerospace, chemical and energy industries around Pau (Nouvelle-Aquitaine).

\textbf{Instrument validity.} Our identification strategy rests on the relevance of lightning strikes in predicting the rollout of fast broadband internet, which in turn is a fundamental pre-condition for the use of cloud technologies. Nevertheless, the exogeneity of lightning strikes may not be sufficient to uphold the exclusion restriction and validity of our identification strategy. One could hypothesise that the association with cloud is driven not necessarily by lightning strike density, but by other unobserved and correlated phenomena. To support our argument, we augment all IV specifications, including ones related to validity checks in Table \ref{tab:ivalidity}, with a battery of controls measured at the level of municipalities. 

First, among these, we include variables capturing geographical features that may correlate with the incidence of lightning strikes. In particular, our instrument tends to take higher values in regions located at higher altitudes, such as the Alps and the Pyrenees, where lightning strikes occur more frequently (see Figure \ref{fig:lightningpc_figure}). Furthermore, terrain ruggedness may act as a barrier to infrastructure development. Therefore, we control for both altitude and terrain ruggedness to mitigate arguments that our instrumental variable is capturing geographical factors correlated with the incidence of lightning strikes. In particular, we include  average elevation of the municipality\footnote{Elevation data is sourced from \href{https://www.usgs.gov/centers/eros/science/usgs-eros-archive-digital-elevation-global-30-arc-second-elevation-gtopo30}{GTOPO30}, a dataset resulting from a collaborative effort led by the staff at the U.S. Geological Survey's EROS Data Center. The data comes as a raster with horizontal grid spacing of 30-arc seconds. Average municipality elevation is obtained by averaging out the vertical units of the elevation raster -- representing elevation in meters above mean sea level -- within each municipality.}.  We also control for the average terrain ruggedness, which is intended to separate the "digging trenches is hard" channel from the energy spike mechanism underlying our instrument.\footnote{Terrain ruggedness is measured using the same data used to measure average elevation of municipality -- GTOPO30. Terrain Ruggedness Index (TRI) \citep{riley1999} quantifies topographic heterogeneity by measuring elevation differences between a focal cell and its immediate neighbours; for each cell in the GTOPO30, TRI is computed as square root of the sum of the squares of the elevation differences between the center cell and each of its 8 neighbours. We average the TRI of the cells contained in each postcode to obtain a postcode-level measure of terrain ruggedness.}  This set of geographical controls is consistently included across all endogenous treatment and two-stage least squares specifications presented in the paper. Adding this set of additional controls leaves the significance of lightning unchanged and indicates that, despite the correlations among these variables, the 'good' variation in broadband connection and cloud use is captured by the lightning variable.

Second, a parallel threat to the instrument’s validity arises from a potential correlation between lightning frequency and local economic development, which could generate technological or growth spillovers affecting our estimates. On the one hand, tall buildings in high-income urban areas are more likely to attract lightning strikes \citep{golde1978lightning, eriksson1987incidence}. On the other hand, if lightning strikes are more frequent in areas with weaker economic activity, the instrument could be spuriously correlated with the outcome variable. To address these concerns, we augment the specification in Table \ref{tab:ivalidity} with municipality-level measures of economic activity and performance -- namely, total employment, average firm growth over the same period as the dependent variable (between $t$ and $t-5$), average firm productivity, and the number of firms. Each measure is computed at the time of cloud use and excludes the firms in our dataset to minimise mechanical correlation. In addition, we control for the rural/urban status of the municipality in which the firm is located\footnote{We employ the definition of degree of urbanisation DEGURBA (\url{https://ec.europa.eu/eurostat/web/degree-of-urbanisation/information-data}), provided by the European Commission, to identify rural municipalities. DEGURBA is based on a set of criteria including population density, size, and contiguity. Spatial units are classified as (i) cities or densely populated areas, (ii) towns and semi-dense areas or intermediate density areas and (iii) rural areas or thinly populated areas. The latter identifies our rural municipalities. For more details on the DEGURBA methodology, visit \url{https://ec.europa.eu/eurostat/statistics-explained/index.php?title=Applying_the_degree_of_urbanisation_manual}.}. The results remain unchanged: lightning frequency continues to predict lower broadband speed. Similarly to elevation and rural status, all specifications reported in the paper include this additional set of variables.

Finally, it is noteworthy to mention that --  as documented by \citep{Andersen2012} -- the economic effect of lightning strikes has only manifested after the beginning of the diffusion of ICTs and internet, as the incidence of this random natural phenomenon has discouraged investments in ICT equipments due to the additional costs imposed on infrastructure management. Moreover, given the stationary nature of this phenomenon (see Table \ref{tab:stationarity_tests}), it is unlikely that other natural events such as climate shocks could account for the reduced form correlation between lightning and firm growth. This provides further support of the validity of the exclusion restriction of our instrument's relationship with firm growth. Also, natural phenomena are not likely to align with administrative boundaries, where firm-specific economic policies are implemented. This rules out the concern that lightning-prone areas systematically differ in terms of regulatory environments that are directly related to firms' growth. It is also unlikely that firm-level growth determinants -- such as managerial capabilities, industry composition, or business cycles -- are directly affected.

\begin{table}[!ht]\centering
\begin{threeparttable}
\scalebox{0.65}{
        \begin{minipage}{\textwidth}
        \centering
            \begin{tabular}{lcccc}
            \toprule
            Test Name & Test Statistic & Lag order & P-value \\
            \midrule
            Augmented Dickey-Fuller (ADF) & -6.12 & 5 & 0.01 \\
            Kwiatkowski-Phillips-Schmidt-Shin (KPSS) & 0.30 & 4 & 0.11 \\
            Phillips-Perron (PP) & -52.59 & 4 & 0.01 \\
            \bottomrule
            \end{tabular}
            \begin{tablenotes}
              \scriptsize
              \item \textit{Note}: The table reports the results of three different stationarity tests conducted on the time series of lightning strike density (average number of daily lightning strikes per square kilometre) in France. Lightning strike density is obtained from the WWLLN dataset \citep{Kaplan2023} and is measured monthly between 2010 and 2020. The null hypotheses of the ADF and PP tests states that the time series is non-stationary, while for the KPSS the null hypothesis states that the series is stationary. 
              \end{tablenotes}
        \end{minipage}}
    \end{threeparttable}
    \caption{Lightning strikes in density -- stationarity tests}\label{tab:stationarity_tests}
\end{table}

To support the validity of our IV, we present several complementary approaches in Table \ref{tab:ivalidity}. First, we assess the validity of our instrumental variable by examining the relationships between broadband speed and cloud (Column 1, Table \ref{tab:ivalidity}) and between lightning strike density and broadband speed (Column 2, Table \ref{tab:ivalidity}). Broadband speed is measured on a categorical scale from 0 to 5 -- corresponding respectively to speeds below 2, between 2 and 10, 10 and 30, 30 and 100, and above 100 Mbit/s. As expected, cloud and broadband connection are significantly and positively associated, confirming that faster broadband connections facilitate the use of cloud. Furthermore, lightning strike density is negatively and significantly associated with broadband speed, consistent with the idea that higher lightning incidence deteriorates connection quality. These relationships remain robust after controlling for a rich set of geographical and firm-level characteristics, reinforcing the validity of our instrument.

\begin{table}[!ht]\centering
\begin{threeparttable}
    \scalebox{0.65}{
        \begin{minipage}{\textwidth}
        \centering
            \begin{tabular}{lccccc}
            \toprule
            & \multicolumn{1}{c}{Probit} & \multicolumn{1}{c}{Ordered Probit} & \multicolumn{2}{c}{Probit} & \multicolumn{1}{c}{OLS} \\
            \cmidrule(lr){2-2} \cmidrule(lr){3-3} \cmidrule(lr){4-5} \cmidrule(lr){6-6}
            & (1) & (2) & (3) & (4) & (5) \\
            & Cloud  & Broadband Speed & E-commerce & Big Data & ICT Workers \\
            \midrule
            Log-Lightning Strike Density &  & $-$0.0363*** & $-$0.0012 & $-$0.0038 & $-$0.0005 \\
            &  & (0.0120)     & (0.0036)  & (0.0037)  & (0.0005) \\\addlinespace
            Broadband Speed: \textless{}2 Mbit/s       & 0.0569**  &            &            &            &            \\
            & (0.0228)  &            &            &            &            \\\addlinespace
            Broadband Speed: between 2 and 10 Mbit/s   & 0.0874*** &            &            &            &            \\
            & (0.0187)  &            &            &            &            \\\addlinespace
            Broadband Speed: between 10 and 30 Mbit/s  & 0.1207*** &            &            &            &            \\
            & (0.0190)  &            &            &            &            \\\addlinespace
            Broadband Speed: between 30 and 100 Mbit/s & 0.1463*** &            &            &            &            \\
            & (0.0193)  &            &            &            &            \\\addlinespace
            Broadband Speed: \textgreater{}100 Mbit/s  & 0.1807*** &            &            &            &            \\
            & (0.0198)  &            &            &            &            \\\addlinespace
            \midrule\addlinespace
            Observations              & 16,216 & 16,216 & 16,216 & 16,216 & 16,216 \\
            Industry, Region, Year FE & X      & X      & X      & X      & X      \\
            Controls                  & X      & X      & X      & X      & X      \\
            Geographic Controls       & X      & X      & X      & X      & X      \\
            Pseudo R$^2$              & 0.176  &        & 0.190  & 0.103  &        \\
            Adj. R$^2$                &        &        &        &        & 0.881  \\
            \bottomrule
            \end{tabular}
            \begin{tablenotes}
                \item \textit{Note}: Column 1 reports average marginal effects from a probit estimation where the dichotomous dependent variable is cloud use. Column 2 reports ordered probit estimates using broadband connection speed at time $t$ as the dependent variable. Columns 3 and 4 report average marginal effects from probit estimations where the dichotomous dependent variables are e-commerce participation and the use of big data analytics, respectively. Column 5 reports OLS estimates using the share of ICT workers in a firm as the dependent variable. Log lightning density is the logarithm of the (population-weighted) density of daily lightning strikes at the municipality level, averaged between 2008 and 2017. Broadband speed is a categorical variable ranging from 0 to 4. It takes values 0, 1, 2, 3, and 4 if the connection speed is below 2, between 2 and 10, between 10 and 30, between 30 and 100, or above 100 Mbit/s. Firm-level controls, measured at $t-5$, include (in logarithmic terms) sales, physical and intangible capital, firm age, and the average hourly wage of managers and engineers, as well as ICT and R\&D shares of hours worked, and dummies for exporting activity and multi-establishment status. Geographic controls, measured at time $t$ at the municipality level, include average firm growth over the same period as the dependent variable, average productivity, total employment, and the number of firms, as well as dummies for elevation and rural areas. All regressions include 2-digit industry, region, and year fixed effects. Standard errors are clustered at the NACE 2-digit level in Columns 2 and 5, and computed via the delta method in Columns 1, 3, and 4. In Column 2, all cut points of the ordered probit estimation are statistically significant at $p < 0.05$. Estimated coefficients of firm-level and geographic controls are not reported but are available upon request. *** $p<0.01$, ** $p<0.05$, * $p<0.1$.
            \end{tablenotes}
        \end{minipage}
    }
\end{threeparttable}
\caption{Instrument validity}
\label{tab:ivalidity}
\end{table}

Second, we assess whether lightning strike density is correlated with firm-level digitalisation. A positive correlation could confound the relationship between lightning and cloud use. Also, as more digitalised firms may exhibit higher growth rates independently of cloud and broadband quality, evaluating this correlation allows us to rule out potential endogeneity sources in our identification strategy. To this end, we perform a placebo test in which the relevance of our instrument is evaluated based on its ability to predict participation in e-commerce activities and the use of big data analytics (Columns 2 and 3, Table \ref{tab:ivalidity}). Similarly, we test whether lightning strike density is correlated with a proxy for firms’ digital intensity -- the share of ICT workers within the firm (Column 4, Table \ref{tab:ivalidity}). Results show that lightning strike density does not predict e-commerce participation, big data analytics use, or ICT intensity. This suggests the absence of indirect relationships between lightning and general digitalisation levels.

Third, we examine whether the transition from copper-based DSL to fibre infrastructure affects the validity of our first stage. By 2016-2018, fibre rollout in France was already substantial, and its susceptibility to lightning-related disruptions may differ from that of copper networks. A potential concern is that areas with higher lightning exposure may have been prioritised for more resilient infrastructure upgrades, including fibre deployment, which could attenuate or even offset the first-stage relationship. To address this issue, we collect municipality-level data on fibre penetration -- measured as the share of buildings eligible for fibre -- from ARCEP for 2016 and 2018.\footnote{Data prior to 2017 are not publicly available and were provided by the data provider ARCEP upon request.} We use these data to assess whether the relationship between lightning density and cloud use depends on the local broadband technology mix. Specifically, we test whether fibre deployment is systematically correlated with lightning density. As shown in Table \ref{tab:fibre}, we find no statistically significant relationship between lightning density and fibre penetration at the municipality level in either year. This suggests that fibre rollout is not disproportionately targeted toward high-lightning areas, alleviating concerns that infrastructure upgrades may weaken or reverse the first-stage relationship. \rev{Furthermore, we assess whether the first-stage relationship varies with fiber penetration by interacting lightning density with fiber penetration bins.\footnote{The excluded bin is fiber penetration below the 50th percentile of the yearly fiber distribution, equal to 0.04\% in 2016 and 32.76\% in 2018.} The results in Table \ref{apdx_tab:intIVFP} show that the negative relationship between lightning density and cloud use persists across fiber penetration groups, with interaction terms that are generally not statistically significant. The one exception is a negative and significant interaction term at the very top of the fiber penetration distribution, driven by firms located in areas above the 99th percentile of the fiber penetration distribution, which however does not constitute a systematic pattern across the distribution and reflects a very small subset of firms. A plausible explanation is that even in high-fiber areas, the associated active equipment and last-mile connections remain vulnerable to power surges induced by lightning strikes, so that reliability disruptions continue to affect cloud use decisions. Moreover, areas with the highest-high fiber penetration are likely to be the same ones that were the first to be fully covered by older infrastructure, making it hard to distinguish the presence of fiber for the exposure of these areas to lightning in previous years, when copper/DSL was still prevalent. Either way, to prove that our results are not driven by this small subset of highly connected municipalities, we re-estimate our TSLS model excluding municipalities above the top 99th quantile of fiber penetration. The results, reported in Table \ref{apdx_tab:IV99FP}, remain unchanged.}

\begin{table}[!htbp] \centering 
    \begin{threeparttable}
        \scalebox{0.65}{
            \begin{minipage}{\textwidth}
					\begin{tabular}{lccccc}
						\toprule
						& (1)      & (2)     & (3)     & (4) & (5)     \\
						\addlinespace\cmidrule(lr){2-6}\addlinespace
						Log-Lightning Strikes Density   & -0.239   & -0.238  & -0.117  & -0.066 & -0.066  \\
						& (0.154)  & (0.154) & (0.082) & (0.054) & (0.047) \\
						\addlinespace\midrule\addlinespace
						Num.Obs.                        & 69504    & 69504   & 69504   & 69504  & 69504   \\
						R2 Adj.                         & 0.001    & 0.031   & 0.097   & 0.099 & 0.099   \\
						Year FE                        &          & X       & X       & X   & X       \\
						Region FE                      &          &         & X       & X     & X       \\
						Elevation, Ruggedness &         &       &  & X & X   \\\bottomrule   
					\end{tabular}
                \begin{tablenotes}
						\item \textit{Note}: The dependent variable is fiber penetration. Observations are French municipality-year (2016-2018). Standard errors are clustered at the French department level in Columns 1-4 and at the municipality level in Column 5. *** p$<$0.01, ** p$<$0.05, * p$<$0.1.                 \end{tablenotes}
            \end{minipage}
        }
    \end{threeparttable}
    \caption{Fibre penetration and lightning strikes}
    \label{tab:fibre}
\end{table}

\begin{table}[!htbp] \centering 
        \begin{threeparttable}
            \scalebox{0.7}{
                \begin{minipage}{\textwidth}
\begin{tabular}{lcccccc}
\toprule
& (1) & (2) & (3) & (4) & (5) & (6) \\
\addlinespace\cmidrule(lr){2-7}\addlinespace
Log-Lightning   Strikes Density                      & -0.0170***           &                      & -0.0136***           & -0.0090*             & -0.0088*             & -0.0086*                   \\
& (0.0033)             &                      & (0.0044)             & (0.0048)             & (0.0049)             & (0.0049)                   \\\addlinespace
50th-60th FP                                         &                      & 0.0162               & 0.0105               & 0.1490               & 0.1484               & 0.1471                     \\
&                      & (0.0126)             & (0.0126)             & (0.1053)             & (0.1057)             & (0.1059)                   \\\addlinespace
60th-70th FP                                         &                      & 0.0057               & -0.0032              & 0.0492               & 0.0474               & 0.0466                     \\
&                      & (0.0111)             & (0.0118)             & (0.1033)             & (0.1037)             & (0.1038)                   \\\addlinespace
70th-80th FP                                         &                      & 0.0353***            & 0.0273*              & 0.0736               & 0.0719               & 0.0701                     \\
&                      & (0.0132)             & (0.0138)             & (0.0766)             & (0.0765)             & (0.0765)                   \\\addlinespace
80th-90th FP                                         &                      & 0.0512***            & 0.0372*              & 0.1206*              & 0.1196*              & 0.1179*                    \\
&                      & (0.0167)             & (0.0193)             & (0.0686)             & (0.0689)             & (0.0689)                   \\\addlinespace
Above 90th FP                                        &                      & 0.0494***            & 0.0349**             & 0.1609**             &  &                            \\
&                      & (0.0137)             & (0.0156)             & (0.0626)             &  &                            \\\addlinespace
90th-95th FP                                         &                      &                      &                      &                      & 0.1483               & 0.1473                     \\
&                      &                      &                      &                      & (0.1396)             & (0.1394)                   \\\addlinespace
Above 95th FP                                        &                      &                      &                      &                      & 0.2076***            &                            \\
&                      &                      &                      &                      & (0.0739)             &                            \\\addlinespace
95th-99th FP                                         &  &  &  &  &  & 0.1590*                    \\
&  &  &  &  &  & (0.0881)                   \\\addlinespace
Above 99th FP                                        &  &  &  &  &  & 0.3949***                  \\
&  &  &  &  &  & (0.1393)                   \\\addlinespace
\multicolumn{2}{l}{50th-60th FP$\times$Log-Lightning Strikes   Density}     &                      &                      & -0.0146              & -0.0145              & -0.0144                    \\
&                      &                      &                      & (0.0111)             & (0.0111)             & (0.0112)                   \\\addlinespace
60th-70th FP$\times$Log-Lightning Strikes   Density  &                      &                      &                      & -0.0050              & -0.0048              & -0.0047                    \\
&                      &                      &                      & (0.0115)             & (0.0116)             & (0.0116)                   \\\addlinespace
70th-80th FP$\times$Log-Lightning Strikes   Density  &                      &                      &                      & -0.0042              & -0.0041              & -0.0039                    \\
&                      &                      &                      & (0.0083)             & (0.0083)             & (0.0083)                   \\\addlinespace
80th-90th FP$\times$Log-Lightning Strikes   Density  &                      &                      &                      & -0.0092              & -0.0091              & -0.0089                    \\
&                      &                      &                      & (0.0088)             & (0.0088)             & (0.0088)                   \\\addlinespace
Above 90th FP$\times$Log-Lightning   Strikes Density &                      &                      &                      & -0.0151*             &                      &                            \\
&                      &                      &                      & (0.0083)             &                      &                            \\\addlinespace
90th-95th FP$\times$Log-Lightning Strikes   Density  &                      &                      &                      &                      & -0.0145              & -0.0144                    \\
&                      &                      &                      &                      & (0.0216)             & (0.0216)                   \\\addlinespace
Above 95th FP$\times$Log-Lightning   Strikes Density &                      &                      &                      &  & -0.0201**            &                            \\
&                      &                      &                      &  & (0.0094)             &                            \\\addlinespace
95th-99th FP$\times$Log-Lightning Strikes   Density  &  &  &  &  &  & -0.0127                    \\
&  &  &  &  &  & (0.0111)                   \\\addlinespace
Above 99th FP$\times$Log-Lightning   Strikes Density &  &  &  &  &  & -0.0494**                 \\
&  &  &  &  &  & (0.0187)                   \\\addlinespace\midrule\addlinespace
Observations                                         & 16,216               & 16,216               & 16,216               & 16,216               & 16,216               & \multicolumn{1}{c}{16,216} \\
Adj R2                                               & 0.194                & 0.194                & 0.194                & 0.194                & 0.194                & \multicolumn{1}{c}{0.194}  \\
Industry, Reg., Year FE                              & X                    & X                    & X                    & X                    & X                    & X      \\
Firm controls                                        & X                    & X                    & X                    & X                    & X                    & X      \\
Municipality controls                                & X                    & X                    & X                    & X                    & X                    & X      \\
Delta                                                &                      &                      &                      & -0.434               & -0.429               & \multicolumn{1}{c}{-0.426} \\\bottomrule
\end{tabular}
                    \begin{tablenotes}
                        \item \textit{Note}: The dependent variable is cloud. FP stands for fiber penetration. Delta is the differente between the average IV of the last category of FP and the rest (e.g, the average IV of firms above the 90th percentile of FP is 43.4\% lower than the one of firms below it). Standard errors are clustered at the NACE 2-digit level. *** p$<$0.01, ** p$<$0.05, * p$<$0.1. 
                    \end{tablenotes}
                \end{minipage}
            }
        \end{threeparttable}
    \caption{Fiber penetration-lightning strikes density interaction}
    \label{apdx_tab:intIVFP}
    \end{table}

    \begin{table}[!htbp] \centering
\begin{threeparttable}
\scalebox{0.7}{
\begin{minipage}{\textwidth}
\begin{tabular}{lccc}
\toprule
& (1)                            & (2)                          & (3)                            \\
& \multicolumn{3}{c}{TSLS without top 99\% fast broadband penetration}                           \\
& 2nd - Growth                   & 1st - Cloud                  & 1st - Cloud $\times$ Log Sales \\\addlinespace\cmidrule(lr){2-4}\addlinespace
Cloud (t)                                              & {1.3162**}   &          &            \\
& {(0.5004)}   &          &            \\\addlinespace
Cloud (t)$\times$Log Sales (t-5)                       & {-0.0910***} &          &            \\
& {(0.0335)}   &          &            \\\addlinespace
Log Sales (t-5)                                        & {-0.0299*}   & {0.0282**} & {0.1994}     \\
& {(0.0155)}   & {(0.0122)} & {(0.1301)}   \\\addlinespace
Log-Lightning Strikes Density                          &            & {0.0020}   & {0.3733***}  \\
&            & {(0.0105)} & {(0.1007)}   \\\addlinespace
Log-Lightning Strikes Density$\times$Log   Sales (t-5) &            & {-0.0020*} & {-0.0598***} \\
&            & {(0.0012)} & {(0.0122)}   \\
\addlinespace\midrule\addlinespace
Observations                                           & 16,041                         & 16,041                       & 16,041                         \\
Firm Controls                                          & X                              & X                            & X                              \\
Municipality Controls                                  & X                              & X                            & X                              \\
Industry, Reg., Year FE                                            & X                              & X                            & X                              \\
KP F Statistic                                         & 11.31                          & 11.31                        & 11.31                          \\
Anderson-Rubin Test                                    & 0.0493                         & 0.0493                       & 0.0493     \\\bottomrule                   
\end{tabular}

\begin{tablenotes}
\small
\item \textit{Note}: The table reports TSLS estimates when excluding firms in the top 99\% of the fiber penetration distribution. Standard errors are clustered at the strata level and reported in parentheses. *** $p<0.01$, ** $p<0.05$, * $p<0.1$.
\end{tablenotes}

\end{minipage}
}
\end{threeparttable}
    \caption{IV Results - Excluding top 99\% fiber penetration}
    \label{apdx_tab:IV99FP}
\end{table}

Taken together, the results in Tables \ref{tab:ivalidity}, \ref{tab:fibre}, \ref{apdx_tab:intIVFP} and \ref{apdx_tab:IV99FP} imply that the IT infrastructure most affected by lightning strikes is specifically the one that is required by fast broadband internet. Furthermore, these results support the hypothesis that lightning strikes influence cloud use only through their impact on broadband availability, and not through broader digitalisation channels.

\textbf{Endogenous Treatment specification}. Despite being widely employed in applied economics, traditional Two-Stages Least Squares estimator present a number of limitations in our empirical setting, mainly imposed by the presence of an endogenous dichotomous variable (cloud use) interacted with a continuous variable (firm size) \citep{wooldridgeCF}. Therefore, we resort to an endogenous treatment regression framework \citep{heckman1976, heckman1978, maddala1983,wooldridgeCF}, a latent variable approach widely used in research \citep{shaver, king, campa} and closely related to conventional instrumental variable models such as two stage least squares \citep{vellaET}. This model allows to address endogeneity issues, such as self-selection into treatment, by simultaneously estimating a selection and an outcome model via Maximum Likelihood Estimation (MLE).

The endogenous treatment method offers several advantages. To start with, it employs a Probit model for the selection equation. This is particularly handy in our case, where the endogenous variable -- cloud use -- is dichotomous. On the technical side, endogenous treatment does not generate predicted values outside the unity range of the probability space, unlike Linear Probability Models (LPM) based on OLS  \citep{hamilton,clougherty}. In addition, the diffusion of cloud technologies is likely to follow an S-curve pattern \citep[as with most technologies, see][]{geroski}, where the use probability accelerates after a slow start, then saturates. This non-linear process is better captured by the Probit specification, in line with the so-called rank models \citep{davies1979diffusion, camerani2016daps} widely used to explain technological use based on individual characteristics. Conversely, the linear nature of the first stage of the TSLS model is less suited to reflect such dynamics. 
Finally, the endogenous treatment model belongs to the family of control function approaches. As such, it allows for the introduction of interaction terms between the endogenous treatment and other variables in a more parsimonious manner compared to the standard TSLS procedure \citep[see the discussion in][]{wooldridgeCF}, which employs the interactions between the exogenous variable and the interacted variable as instruments. \rev{However, this comes at the cost of assuming joint normality of the errors in the selection and outcome equations. This is not an innocuous assumption in our context, because firm growth rates are well known to be fat-tailed \citep{bosemodel}, and departures from normality could in principle affect our estimates. As a check that results in Table \ref{tab:IV} do not rely on this assumption, Table \ref{tab:etbscheck} estimates a specification with bootstrapped standard errors, which yields estimates consistent with the ET results reported in the main text.}

\begin{table}[!htbp] \centering
    \begin{threeparttable}
        \scalebox{0.65}{\color{black}
            \begin{minipage}{\textwidth}
            \begin{tabular}{lcc}
            \toprule
                                      & (1)        & (2)        \\
                                      \addlinespace\cmidrule(lr){2-3}\addlinespace
Cloud (t)                        & 0.2885***     &    \\
                                      & (0.0587)   &    \\\addlinespace
Cloud (t)$\times$Log Sales (t-5) & -0.0170***    &     \\
                                      & (0.0049)   &    \\\addlinespace
Log Sales (t-5)                       & -0.0421*** & -0.1546*** \\
                                      & (0.0052)   & (0.0112)   \\
                                      \addlinespace
Log-Lightning Strikes Density                       &  & -0.0571*** \\
                                      &   & (0.0140)   \\
                                      \addlinespace\midrule\addlinespace
Observations                          & 16,216      & 16,216     \\
Industry, Reg., Year FE               & X          & X          \\
Controls                              & X          & X         \\
\bottomrule
                \end{tabular}
                \begin{tablenotes}
                    \small
                    \item \textit{Note}: The table reports the results of the Endogenous Treatment estimation of Equation \ref{eq:empirical_model_ET}. Column 1 and 2 report the results of the outcome and selection equations, respectively. Controls are measured at $t-5$ and include the logarithms of age, tangible capital, the share of workers in ICT and R\&D occupations, the average hourly wage of managers and engineers, and two dummies for exporter status and multi-establishment firms. Standard errors are based on 1000 bootstrap replications. The full version of this table, including coefficients for all controls, is available upon request. *** $p<0.01$, ** $p<0.05$, * $p<0.1$.   
                \end{tablenotes}
            \end{minipage}
        }
    \end{threeparttable}
    \caption{Endogenous Treatment Estimation with Bootstrapped Standard Errors}
    \label{tab:etbscheck}
\end{table}

In the context of the endogenous treatment framework, Equation \ref{eq:baseline_OLS} reads as follows:

\begin{gather}\label{eq:empirical_model_ET}
	\begin{aligned}
		& \text{Sales Growth Rate}_{i, t, t-5} = \alpha + \beta_1 \text{Cloud}_{i,t} +  \beta_2 \text{Cloud}_{i,t} \cdot \text{Log-Sales}_{i,t-5} + \mathbf{\beta_X} \mathbf{X}_{i,t-5} \\
          & + \text{2-digit Ind.}_{j} + \text{Region}_{r} + \text{Year}_{t} + \epsilon_{i,t}\\		
		& \text{Cloud}_{i,t} =
		\begin{cases}
			1, & \text{if}\ \ \beta_Z \text{Log-Lightning Density}_r + \mathbf{\beta_X} \mathbf{X}_{i,t-5} + \omega_{i,t} > 0 \\
			0, & \text{otherwise} 
		\end{cases}\\
	\end{aligned}
\end{gather}

\medskip 

\noindent where $\text{Cloud}_{i,t}$ is the endogenous dummy variable for the use of cloud, $\mathbf{X}_{i,t-5}$ is a vector of controls including the same variables of Equation \ref{eq:baseline_OLS} and the variable $\text{Log-Lightning Density}_r$, which is excluded from the outcome equation. The estimation of the endogenous treatment model employs the Full Information Maximum Likelihood (FIML) method, concurrently estimating the selection equation (using the dummy variable for cloud use as the dependent variable) and the outcome equation (i.e., the sales growth rate regression).\footnote{This process assumes the joint normality of errors $(\epsilon_i, \omega_i)$.} 

The endogenous treatment model produces estimates robust to the presence of specification errors when an additional variable is incorporated into the selection equation, but omitted from the outcome equation. This variable must adhere to two conditions \citep{puhani2000}, similarly to the standard IV assumptions in TSLS models: it must be strongly correlated with the endogenous dummy variable (relevance condition) and exogenous with respect to the outcome equation (exclusion restriction), conditional on the included controls.

\textbf{Two Stage Least Square specification}. In addition to estimating an endogenous treatment model, we also resort to a traditional TSLS estimator. Indeed, the assumption of joint normality of errors may be problematic when considering growth rates, whose distribution is typically fat-tailed \citep{bosemodel}. The estimation is complicated by the presence of an interacted endogenous variable (Equation \ref{eq:baseline_OLS}), and only one instrument. For this specific case, we use a non-linear TSLS regression incorporating the interaction between cloud and sales as additional endogenous variable, instrumented using the interaction between sales and lightning strike density \citep{wooldridgeCF}.\footnote{It is worth mentioning that TSLS models are generally more consistent but less efficient than control function approaches in presence of explanatory variables that are interacted with the instrumented endogenous variable.} We therefore instrument $Cloud_{i,t}$ and its interaction with initial size using the log-density of lightning strikes in the firm's municipality, and its interaction with log sales. The identifying assumption is that lightning strike density affects firm growth only through its effect on cloud use. 
\rev{This assumption for the interaction instrument requires that lightning density not affect firm growth differentially by size through channels other than cloud use. A natural concern is that lightning-prone areas have historically had thinner infrastructure more broadly, a constraint that could bind more tightly on smaller, locally-oriented firms than on larger, nationally-active firms with access to infrastructure elsewhere. We view this concern as attenuated by our specification. Municipality-level controls -- rural status, elevation, terrain ruggedness, and local economic conditions (average growth, productivity, employment, and firm counts) -- plausibly absorb these component of local infrastructure that could otherwise confound the interaction term.}

We need two separate first stage regressions for each endogenous regressor in Equation \ref{eq:baseline_OLS}. The first stage for cloud use ($\text{Cloud}_{i,t}$) is given by the following equation:

\begin{equation}\label{eq:first_stage_cloud}
\begin{split}
    & \text{Cloud}_{i,t} = \\
    & \alpha_1 \ln \text{Log-Sales}_{i,t-5} + \alpha_2 \text{Log-Lightning Density}_r + \\
    & \alpha_3 \text{Log-Lightning Density}_r \times  \text{Log-Sales}_{i,t-5} + \alpha_{\textbf{X}}\mathbf{X}_{i,t-5} + \mu_j + \sigma_r + \tau_t + \nu_{i,t}
\end{split}
\end{equation}

By the same token, the first stage for the interaction of cloud with initial firms' size ($\text{Log-Sales}_{i, t-5}$) reads as follows:

\begin{equation}\label{eq:first_stage_interaction}
\begin{split}
    & \text{Cloud}_{i,t-5} \times \text{Log-Sales}_i = \\
    & \pi_1 \text{Log-Sales}_{i,t-5} + \pi_2 \text{Log-Lightning Density}_r + \\
    & \pi_3 \text{Log-Lightning Density}_r \times \text{Log-Sales}_{i,t-5} + \pi_{\mathbf{X}}\mathbf{X}_{i,t-5} + \mu_j + \sigma_r + \tau_t + \eta_{i,t}
\end{split}
\end{equation}

\noindent The two excluded instruments defined as dependent variables of Equations \ref{eq:first_stage_cloud} and \ref{eq:first_stage_interaction} identify the two endogenous regressors of the second stage regressions, yielding an exactly identified model:

\begin{equation}\label{eq:second_stage}
\begin{split}
    & \text{Sales Growth}_{i,t,t-5} = \\
    & \beta_0 + \beta_1 \widehat{\text{Cloud}}_{i,t} + \beta_2 \widehat{\text{Cloud}_{i,t} \times \text{Log-Sales}}_{i,t-5} + \\
    & \beta_3 \text{Log-Sales}_{i,t-5} + \beta_{\textbf{X}}\mathbf{X}_{i,t-5} + \text{2-digit Ind.}_j + \text{Region}_r + \text{Year}_t + \varepsilon_{i,t}
\end{split}
\end{equation}

\noindent where $\text{Sales Growth}_{i,t,t-5}$ denotes the sales growth rate of firm $i$ between $t-5$ and $t$. The interaction between cloud use and initial size allows the effect of cloud computing to vary across the firm size distribution. $\mathbf{X}_{i,t-5}$ is a vector of firm- and municipality-level controls measured at $t-5$. Firm-level controls are the same of Equation \ref{eq:baseline_OLS}. Municipality-level controls include a dummy for rural status, average elevation, average terrain ruggedness, average firm growth, average firm productivity, total number of firms, and total employment. Fixed effects remain unchanged from Equation \ref{eq:baseline_OLS}. We assess the strength of the instruments through the Kleibergen-Paap F statistic and report Anderson-Rubin test $p$-values for inference that is robust to weak instruments.

\newpage
\clearpage

\section{Additional Tables and Figures}
\label{apdx_sec:full_tables}

\setcounter{figure}{0}
\renewcommand{\thefigure}{B\arabic{figure}}

\setcounter{table}{0}
\renewcommand{\thetable}{B\arabic{table}}
\renewcommand{\theHtable}{table.B\arabic{table}} 

\begin{table}[!htbp]\centering
    \begin{threeparttable}
        \scalebox{0.5}{
            \begin{minipage}{\textwidth}
            \centering
                \begin{tabular}{lccccccc}
                    \toprule
                    & (1) & (2) & (3) & (4) & (5) & (6) & (7) \\
                    & \multicolumn{2}{c}{Endogenous Treatment} & \multicolumn{3}{c}{Two Stage Least Square} & \multicolumn{2}{c}{IMR-augmented} \\
                    & 2nd & 1st & 2nd & 1st & 1st & 2nd & 1st \\
                    & Growth & Cloud & Growth & Cloud & Cloud $\times$ Log Sales & Growth & Cloud \\
                    \addlinespace\cmidrule(lr){2-3}\cmidrule(lr){4-6}\cmidrule(lr){7-8}\addlinespace
                    
                    Cloud (t) 
                    & 0.2885*** &  & 1.3126*** &  &  &  &  \\
                    & (0.0771)  &  & (0.4923)  &  &  &  &  \\\addlinespace
                    
                    Cloud (t)$\times$Log Sales (t-5) 
                    & -0.0170*** &  & -0.0918*** &  &  &  &  \\
                    & (0.0062)   &  & (0.0332)   &  &  &  &  \\\addlinespace
                    
                    Cloud IaaS (t)
                    &  &  &  &  &  & 0.0893 &  \\
                    &  &  &  &  &  & (0.0702) &  \\\addlinespace
                    
                    Cloud SaaS (t)
                    &  &  &  &  &  & 0.3095*** &  \\
                    &  &  &  &  &  & (0.0738)  &  \\\addlinespace
                    
                    Cloud IaaS (t)$\times$Log Sales (t-5)
                    &  &  &  &  &  & -0.0022 &  \\
                    &  &  &  &  &  & (0.0061) &  \\\addlinespace
                    
                    Cloud SaaS (t)$\times$Log Sales (t-5)
                    &  &  &  &  &  & -0.0257*** &  \\
                    &  &  &  &  &  & (0.0065)   &  \\\addlinespace
                    
                    Log Sales (t-5)
                    & -0.0421*** & 0.1546*** & -0.0290* & 0.0276** & 0.1916 & -0.0398*** & 0.1547*** \\
                    & (0.0059)   & (0.0207)  & (0.0149) & (0.0120) & (0.1287) & (0.0049)   & (0.0114)  \\\addlinespace
                    
                    Log-Lightning Strikes Density
                    &  & -0.0571*** &  & 0.0020 & 0.3731*** &  & -0.0568*** \\
                    &  & (0.0130)   &  & (0.0104) & (0.0990) &  & (0.0141)  \\\addlinespace
                    
                    Log-Lightning Strikes Density$\times$Log Sales (t-5)
                    &  &  &  & -0.0020* & -0.0603*** &  &  \\
                    &  &  &  & (0.0012) & (0.0120)   &  &  \\\addlinespace
                    
                    IMR-based generalised error
                    &  &  &  &  &  & 0.0009 &  \\
                    &  &  &  &  &  & (0.0170) &  \\\addlinespace
                    Log-Physical Capital (t-5)                      & 0.0098                   & 0.0077**               & 0.0844**     & 0.0112               & 0.0263**                       & 0.0264**             & 0.0113               \\
                                                & (0.0088)                 & (0.0035)               & (0.0353)     & (0.0082)             & (0.0113)                       & (0.0133)             & (0.0071)             \\\addlinespace
Log-Intangible Capital (t-5)                    & -0.0004                  & 0.0034                 & 0.0730       & -0.0026              & 0.0022                         & 0.0019               & -0.0023              \\
                                                & (0.0090)                 & (0.0056)               & (0.0593)     & (0.0084)             & (0.0189)                       & (0.0137)             & (0.0077)             \\\addlinespace
Log-Age (t-5)                                  & -0.0806***               & -0.0286***             & -0.2848***   & -0.0883***           & -0.0931***                     & -0.0930***           & -0.0885***           \\
                                                & (0.0103)                 & (0.0058)               & (0.0570)     & (0.0111)             & (0.0212)                       & (0.0144)             & (0.0054)             \\\addlinespace
Share of R\&D Hours (t-5)                       & 0.2131***                & 0.0256                 & 0.3469       & 0.2127***            & 0.0488                         & 0.0496               & 0.1779***            \\
                                                & (0.0597)                 & (0.0381)               & (0.3921)     & (0.0519)             & (0.1305)                       & (0.1515)             & (0.0632)             \\\addlinespace
Exporter (t-5)                                  & -0.0081                  & 0.0617***              & 0.5649***    & 0.0132               & 0.2087***                      & 0.2085***            & 0.2096***            \\
                                                & (0.0230)                 & (0.0204)               & (0.2162)     & (0.0124)             & (0.0628)                       & (0.0277)             & (0.0662)             \\\addlinespace
Share of ICT Hours (t-5)                        & 0.0904                   & 0.1912***              & 1.4803***    & 0.1740**             & 0.5443***                      & 0.5441***            & 0.0142               \\
                                                & (0.0994)                 & (0.0548)               & (0.4786)     & (0.0707)             & (0.1713)                       & (0.1343)             & (0.0090)             \\\addlinespace
Log-Average Hourly Wage Managers   (t-5)            & -0.0052                  & 0.0066*                & -0.0380      & 0.0054               & 0.0492***                      & 0.0494***            & 0.0051               \\
                                                & (0.0057)                 & (0.0039)               & (0.0397)     & (0.0033)             & (0.0142)                       & (0.0105)             & (0.0032)             \\\addlinespace
Multiplant (t-5)                                & -0.0238*                 & 0.0381***              & 0.3705***    & -0.0120              & 0.1295***                      & 0.1295***            & -0.0114              \\
                                                & (0.0135)                 & (0.0077)               & (0.0849)     & (0.0087)             & (0.0252)                       & (0.0262)             & (0.0087)             \\\addlinespace
Rural                                           & -0.0108                  & 0.0436***              & 0.3901***    & -0.0006              & 0.1464***                      & 0.1464***            & -0.0006              \\
                                                & (0.0160)                 & (0.0117)               & (0.1200)     & (0.0151)             & (0.0441)                       & (0.0415)             & (0.0111)             \\\addlinespace
Elevation                                       & 0.0436                   & -0.0279                & -0.4831*     & 0.0387               & -0.0903                        & -0.0906              & 0.0366               \\
                                                & (0.0308)                 & (0.0261)               & (0.2567)     & (0.0271)             & (0.1120)                       & (0.1175)             & (0.0279)             \\\addlinespace
Ruggedness                                      & -0.0001                  & 0.0000                 & 0.0007       & -0.0001              & 0.0001                         & 0.0001               & -0.0001              \\
                                                & (0.0001)                 & (0.0001)               & (0.0006)     & (0.0001)             & (0.0002)                       & (0.0002)             & (0.0001)             \\\addlinespace
Log-Total Employment (t)                        & -0.0083                  & 0.0213*                & 0.2018*      & -0.0015              & 0.0854**                       & 0.0855***            & -0.0013              \\
                                                & (0.0109)                 & (0.0109)               & (0.1133)     & (0.0095)             & (0.0403)                       & (0.0299)             & (0.0090)             \\\addlinespace
Average Growth Rate (t)                         & 0.0995***                & -0.0210                & -0.2981      & 0.1001***            & -0.0739                        & -0.0754              & 0.0975***            \\
                                                & (0.0294)                 & (0.0339)               & (0.3458)     & (0.0283)             & (0.1311)                       & (0.1074)             & (0.0291)             \\\addlinespace
Log-Average Productivity (t)                    & -0.0045                  & 0.0558***              & 0.6294***    & 0.0054               & 0.1716***                      & 0.1714***            & 0.0065               \\
                                                & (0.0192)                 & (0.0107)               & (0.1343)     & (0.0141)             & (0.0341)                       & (0.0344)             & (0.0114)             \\\addlinespace
Log-Total Number of Firms                       & 0.0042                   & -0.0161                & -0.1694      & 0.0022               & -0.0683                        & -0.0683**            & 0.0020               \\
                                                & (0.0103)                 & (0.0117)               & (0.1217)     & (0.0107)             & (0.0424)                       & (0.0326)             & (0.0095)             \\\addlinespace
                    \midrule
                    
                    Observations 
                    & 16,216 & 16,216 & 16,216 & 16,216 & 16,216 & 16,216 & 16,216 \\
                    
                    Industry, Reg. Year FE 
                    & X & X & X & X & X & X & X \\
                    
                    KP F Statistic 
                    &  &  & 11.69 & 11.69 & 11.69 &  &  \\
                    
                    Anderson-Rubin Test 
                    &  &  & 0.0416 & 0.0416 & 0.0416 &  &  \\
                    
                    \bottomrule      
                \end{tabular}
                \begin{tablenotes}
                    \small
                    \item \textit{Note}: Columns 1-2 and Columns 3-5 correspond to an Endogenous Treatment and TSLS estimation respectively. Columns 6-7 report IMR-augmented estimates, where the selection equation is estimated via probit and the inverse Mills ratio is included in the outcome equation. Standard errors in endogenous treatment and TSLS specifications and Column 7 are clustered at the 2-digit industry level, they are based on 1000 bootstrap replications in Column 6.  *** p$<$0.01, ** p$<$0.05, * p$<$0.1.
                \end{tablenotes}
            \end{minipage}
            }
        \end{threeparttable}
        \caption{Endogenous Treatment and TSLS results}
        \label{tab:IVfull}
\end{table}

\begin{table}[!htbp] \centering 
		\begin{threeparttable}
			\scalebox{0.65}{
				\begin{minipage}{\textwidth}
					\begin{tabular}{lcccc}
						\toprule
						& (1) & (2) & (3) & (4) \\
						Size quantile & SaaS OLS & Cloud OLS & SaaS IV  & Cloud IV \\
						\addlinespace\cmidrule(lr){2-5}\addlinespace
						1st           & 16,15\%  & 16,89\% & 15,94\%   & 77,64\%  \\
						5th           & 14,27\%  & 15,68\% & 14,07\%   & 70,97\%  \\
						10th          & 13,31\%  & 15,06\% & 13,12\%   & 67,59\%  \\
						25th          & 11,43\%  & 13,85\% & 11,26\%   & 60,92\%  \\
						50th          & 8,08\%   & 11,70\% & 7,94\%    & 49,07\%  \\
						75th          & 2,84\%   & 8,32\%  & 2,75\%    & 30,54\%  \\
						90th          & -0,90\%  & 5,91\%  & -0,96\%   & 17,30\%  \\
						95th          & -2,82\%  & 4,68\%  & -2,87\%   & 10,49\%  \\
						99th          & -6,63\%  & 2,22\%  & -6,64\%   & -2,98\% \\
						\bottomrule
					\end{tabular}
					\begin{tablenotes}
						\vspace{0.1cm}
						\item \textit{Note}: Results of $d\textbf{Growth Rate}_{i,t,t-5} = \frac{\partial \textbf{Growth Rate}_{i,t,t-5}}{\partial \textbf{Cloud}_{i,t}} = \beta_1 + \beta_3 \times \textbf{Log Sales}_{i, t-5}$ for different size percentiles.  Columns 1-2 refer to the OLS results for all cloud (Table \ref{tab:OLS}, Column 4) and SaaS cloud (Table \ref{tab:OLS_types}, Column 1). Columns 3-4 refer to the IV results for all cloud and SaaS cloud (Table \ref{tab:IV}, Columns 2 and 6 respectively).
					\end{tablenotes}
				\end{minipage}
			}
		\end{threeparttable}
        \caption{Effect magnitude}
        \label{tab:effects}
	\end{table}

\begin{table}[!htbp] \centering 
    \begin{threeparttable}
        \scalebox{0.65}{
            \begin{minipage}{\textwidth}
                \begin{tabular}{lcccc}
                    \toprule
                    & (1)           & (2)            & (3)                        & (4)                       \\
                    & Manufacturing & Other Services & Advanced Business Services & Utilities \& Construction \\
                    \addlinespace\cmidrule(lr){2-5}\addlinespace
                    Cloud IaaS (t)                        & 0.1157        & 0.0750         & 0.0745                     & -0.1360                   \\
                    & (0.0978)      & (0.0962)       & (0.1784)                   & (0.1894)                  \\\addlinespace
                    Cloud SaaS (t)                        & 0.2014*       & 0.4221***      & 0.1962                     & 0.2136                    \\
                    & (0.1155)      & (0.1091)       & (0.1866)                   & (0.2189)                  \\\addlinespace
                    Cloud IaaS (t)$\times$Log Sales (t-5) & -0.0055       & -0.0032        & 0.0041                     & 0.0208                    \\
                    & (0.0089)      & (0.0087)       & (0.0180)                   & (0.0192)                  \\\addlinespace
                    Cloud SaaS (t)$\times$Log Sales (t-5) & -0.0190*      & -0.0320***     & -0.0175                    & -0.0203                   \\
                    & (0.0102)      & (0.0095)       & (0.0183)                   & (0.0214)                  \\\addlinespace
                    Log Sales (t-5)                       & -0.0478***    & -0.0277***     & -0.0794***                 & -0.0190                   \\
                    & (0.0096)      & (0.0057)       & (0.0194)                   & (0.0128)                  \\\addlinespace\midrule\addlinespace
                    Observations                          & 4,228         & 7,401          & 2,458                      & 1,823                     \\
                    Industry, Reg., Year FE               & X             & X              & X                          & X                         \\
                    Controls                              & X             & X              & X                          & X                         \\
                    Adj R2                                & 0.0803        & 0.106          & 0.116                      & 0.0609           \\\addlinespace\bottomrule        					
                \end{tabular}
                \begin{tablenotes}
                    \item \textit{Note}: The table reports the results of the estimation of Equation \ref{eq:baseline_OLS}, disaggregated by sector. Manufacturing includes NAF sectors 10–33. Advanced Business Services comprises ICT services (NAF 58–63) and professional, scientific, and technical activities (NAF 69–75). Other Services covers wholesale and retail trade (NAF 45–47), transportation and storage (NAF 49–53), accommodation and food services (NAF 55–56), real estate activities (NAF 68), and administrative and support service activities (NAF 77–82). Finally, Utilities \& Construction includes NAF sectors 35–43. Controls are measured in $t-5$ and include the logarithms of age, tangible and intangible capital, the share of workers specialised in ICT and R\&D roles, the average hourly wage of managers and engineers, and two dummies for exporter and multi establishment status. Standard errors are clustered clustered at NACE 2-digit level. The complete version of this table, including additional coefficients of controls, is available upon request. *** p$<$0.01, ** p$<$0.05, * p$<$0.1.
                \end{tablenotes}
            \end{minipage}
        }
    \end{threeparttable}
    \caption{Sectors}
    \label{tab:sectors}
\end{table}

\newpage
\begin{landscape}

\begin{table}[!htbp]\centering
	\begin{threeparttable}
		\scalebox{0.65}{
				\begin{minipage}{\textwidth}
					\begin{tabular}{lccccccccc}
						\toprule
						& (1)          & (2)          & (3) &  (4)          & (5)          & (6)          & (7)          & (8)          & (9)                  \\
						& \multicolumn{3}{c}{Exclude ICT Services} & \multicolumn{3}{c}{Year $\times$ Reg. FE} & \multicolumn{3}{c}{$t{+}5$;$t{-}5$ Growth} \\
						& 2nd          & 1st          & 1st          & 2nd          & 1st          & 1st          & 2nd          & 1st          & 1st          \\
						& Growth & Cloud & Cloud $\times$ Log Sales & Growth & Cloud & Cloud $\times$ Log Sales & Growth & Cloud & Cloud $\times$ Log Sales \\
						\addlinespace\cmidrule(lr){2-4}\cmidrule(lr){5-7}\cmidrule(lr){8-10}\addlinespace
						Cloud (t)                                                & 1.0941**     &              &              & 1.3023**     &              &              & 1.5358**     &              &              \\
						& (0.5293)     &              &              & (0.4938)     &              &              & (0.7093)     &              &              \\\addlinespace
						Cloud (t)$\times$Log Sales (t-5)                         & -0.0789**    &              &              & -0.0915***   &              &              & -0.1143**    &              &              \\
						& (0.0361)     &              &              & (0.0332)     &              &              & (0.0489)     &              &              \\\addlinespace
						Log Sales (t-5)                                          & -0.0309**    & 0.0203**     & 0.1285       & -0.0288*     & 0.0277**     & 0.1936       & -0.0235      & 0.0306**     & 0.2161       \\
						& (0.0143)     & (0.0103)     & (0.1163)     & (0.0148)     & (0.0119)     & (0.1280)     & (0.0164)     & (0.0125)     & (0.1357)     \\\addlinespace
						Log-Lightning Strikes Density                            &              & 0.0097       & 0.4301***    &              & 0.0019       & 0.3730***    &              & -0.0012      & 0.3485***    \\
						&              & (0.0086)     & (0.0896)     &              & (0.0104)     & (0.0992)     &              & (0.0099)     & (0.0982)     \\\addlinespace
						Log-Lightning Strikes Density $\times$ Log Sales (t-5)   &              & -0.0029***   & -0.0674***   &              & -0.0020*     & -0.0604***   &              & -0.0018      & -0.0579***   \\
						&              & (0.0010)     & (0.0110)     &              & (0.0012)     & (0.0120)     &              & (0.0011)     & (0.0113)     \\
						\addlinespace\midrule\addlinespace
						Observations                                             & 15,781       & 15,781       & 15,781       & 16,216       & 16,216       & 16,216       & 14,016       & 14,016       & 14,016       \\
						Firm Controls                                            & X            & X            & X            & X            & X            & X            & X            & X            & X            \\
						Municipality Controls                                    & X            & X            & X            & X            & X            & X            & X            & X            & X            \\
						Industry FE                                              & X            & X            & X            & X            & X            & X            & X            & X            & X            \\
						Year FE                                                  & X            & X            & X            &              &              &              & X            & X            & X            \\
						Region FE                                                & X            & X            & X            &              &              &              & X            & X            & X            \\
						Year $\times$ Region FE                                  &              &              &              & X            & X            & X            &              &              &              \\
						KP F Statistic                                           & 10.37        & 10.37        & 10.37        & 11.81        & 11.81        & 11.81        & 12.101       & 12.101       & 12.101       \\
						Anderson-Rubin Test                                      & 0.0832       & 0.0832       & 0.0832       & 0.0429       & 0.0429       & 0.0429       & 0.0901       & 0.0901       & 0.0901       \\
						\bottomrule
					\end{tabular}
					\begin{tablenotes}
						\item \textit{Note}: The table reports TSLS estimates across different specifications. Columns 1-3 exclude ICT services, Columns 4-6 include year $\times$ region fixed effects, and Columns 7-9 use alternative growth rate definitions ($t{+}5$, $t{-}5$). Firm controls are measured in $t-5$ and include the logarithms of age, tangible and intangible capital, the share of workers specialised in ICT and R\&D roles, the average hourly wage of managers and engineers, and two dummies for exporter and multi establishment status.  Standard errors are clustered clustered at NACE 2-digit level. Municipality controls include average elevation, average terrain ruggedness, a dummy for rural status, average firm growth, average firm productivity, total number of firms, total employment. The complete version of this table, including additional coefficients of controls, is available upon request. Standard errors are clustered at the 2-digit level. *** p$<$0.01, ** p$<$0.05, * p$<$0.1.
					\end{tablenotes}
				\end{minipage}
		}
	\end{threeparttable}
	\caption{Additional TSLS results}
	\label{tab:robtsls}
\end{table}

\end{landscape}

\newpage

\begin{table}[!htbp]
    \centering
    \begin{threeparttable}
        \scalebox{0.65}{
            \begin{minipage}{\textwidth}
                \begin{tabular}{lcccc}
                    \toprule
                    & (1)              & (2)              & (3)              & (4)              \\
                    & All ICT surveys  & No ICT services  & Employment       & IaaS Types       \\
                    \addlinespace\cmidrule(lr){2-5}\addlinespace
                    Cloud IaaS (t)
                    & 0.1948***        & 0.0938           & 0.0656           &                  \\
                    & (0.0665)         & (0.0684)         & (0.0419)         &                  \\
                    \addlinespace
                    Cloud SaaS (t)
                    & 0.2063***        & 0.2685***        & 0.1489***        & 0.3041***        \\
                    & (0.0693)         & (0.0737)         & (0.0451)         & (0.0846)         \\
                    \addlinespace
                    Cloud IaaS (t)$\times$Log Sales (t-5)
                    & -0.0129**        & -0.0025          &                  &                  \\
                    & (0.0061)         & (0.0063)         &                  &                  \\
                    \addlinespace
                    Cloud SaaS (t)$\times$Log Sales (t-5)
                    & -0.0163**        & -0.0222***       &                  & -0.0252***       \\
                    & (0.0062)         & (0.0060)         &                  & (0.0071)         \\
                    \addlinespace
                    Log Sales (t-5)
                    & -0.0316***       & -0.0406***       &                  & -0.0394***       \\
                    & (0.0079)         & (0.0060)         &                  & (0.0059)         \\
                    \addlinespace
                    Cloud IaaS (t)$\times$Log Employment (t-5)
                    &                  &                  & 0.0004           &                  \\
                    &                  &                  & (0.0078)         &                  \\
                    \addlinespace
                    Cloud SaaS (t)$\times$Log Employment (t-5)
                    &                  &                  & -0.0221***       &                  \\
                    &                  &                  & (0.0073)         &                  \\
                    \addlinespace
                    Log Employment (t-5)
                    &                  &                  & -0.0456***       &                  \\
                    &                  &                  & (0.0066)         &                  \\
                    \addlinespace
                    Cloud IaaS - Storage (t)
                    &                  &                  &                  & 0.0483           \\
                    &                  &                  &                  & (0.0606)         \\
                    \addlinespace
                    Cloud IaaS - Computing (t)
                    &                  &                  &                  & 0.1655           \\
                    &                  &                  &                  & (0.1124)         \\
                    \addlinespace
                    Cloud IaaS - Storage (t)$\times$Log Employment (t-5)
                    &                  &                  &                  & 0.0014           \\
                    &                  &                  &                  & (0.0056)         \\
                    \addlinespace
                    Cloud IaaS - Computing (t)$\times$Log Employment (t-5)
                    &                  &                  &                  & -0.0137          \\
                    &                  &                  &                  & (0.0095)         \\
                    \addlinespace\midrule\addlinespace
                    Observations
                    & 30,496           & 15,781           & 16,047           & 16,216           \\
                    Industry, Reg., Year FE
                    & X                & X                & X                & X                \\
                    Controls
                    & X                & X                & X                & X                \\
                    Adj. R2
                    & 0.129            & 0.0916           & 0.0949           & 0.0989           \\
                    \bottomrule
                \end{tabular}
                \begin{tablenotes}
                    \vspace{0.1cm}
                    \item \textit{Note}: The table reports the results of the estimation of Equation \ref{eq:baseline_OLS}, with variations in sample and variables. Column 1 includes additional ICT survey data (2020-2021). Column 2 excludes the ICT service sector from the sample. Column 3 uses employment-based measures to compute growth rates and initial size instead of sales. Column 4 disaggregates IaaS cloud into storage and computing categories. Manufacturing includes NAF sectors 10-33. Advanced Business Services comprises ICT services (NAF 58-63) and professional, scientific, and technical activities (NAF 69-75). Other Services includes wholesale and retail trade (NAF 45-47), transportation and storage (NAF 49-53), accommodation and food services (NAF 55-56), real estate activities (NAF 68), and administrative and support service activities (NAF 77-82). Utilities and Construction includes NAF sectors 35-43. Controls are measured at $t-5$ and include the logarithms of age, tangible and intangible capital, the share of workers in ICT and R\&D occupations, the average hourly wage of managers and engineers, and two dummies for exporter status and multi-establishment firms. Standard errors are clustered at the NACE 2-digit level. The full version of this table, including coefficients for all controls, is available upon request. *** $p<0.01$, ** $p<0.05$, * $p<0.1$.
                \end{tablenotes}
            \end{minipage}
        }
    \end{threeparttable}
    \caption{Robustness -- Alternative samples and variables definitions}
    \label{tab:robsample}
\end{table}

\newpage

\begin{table}[!ht]\centering
\begin{threeparttable}
    \scalebox{0.65}{
        \begin{minipage}{\textwidth}
        \centering
            \begin{tabular}{lcc}
            \toprule
             & (1) & (2) \\
             & Sales Quartiles & Sales \& Age Quartiles \\
            \addlinespace\cmidrule(lr){2-3}\addlinespace
            Cloud IaaS (t) & 0.0904*** & 0.1065** \\
             & (0.0329) & (0.0403) \\
            \addlinespace
            Cloud SaaS (t) & 0.1562*** & 0.1842*** \\
             & (0.0482) & (0.0578) \\
            \addlinespace
            Cloud IaaS (t)$\times$2nd Quartile Sales & 0.0041 & 0.0016 \\
             & (0.0393) & (0.0412) \\
            \addlinespace
            Cloud IaaS (t)$\times$3rd Quartile Sales & -0.0472 & -0.0519 \\
             & (0.0544) & (0.0564) \\
            \addlinespace
            Cloud IaaS (t)$\times$4th Quartile Sales & -0.0461 & -0.0524 \\
             & (0.0367) & (0.0431) \\
            \addlinespace
            Cloud SaaS (t)$\times$2nd Quartile Sales & -0.1213** & -0.0991** \\
             & (0.0479) & (0.0490) \\
            \addlinespace
            Cloud SaaS (t)$\times$3rd Quartile Sales & -0.1092*** & -0.0744* \\
             & (0.0392) & (0.0408) \\
            \addlinespace
            Cloud SaaS (t)$\times$4th Quartile Sales & -0.1740*** & -0.1292*** \\
             & (0.0459) & (0.0440) \\
            \addlinespace
            2nd Quartile Sales & -0.0845*** & -0.0870*** \\
             & (0.0129) & (0.0124) \\
            \addlinespace
            3rd Quartile Sales & -0.0644*** & -0.0680*** \\
             & (0.0178) & (0.0170) \\
            \addlinespace
            4th Quartile Sales & -0.0614** & -0.0641** \\
             & (0.0273) & (0.0275) \\
            \addlinespace
            2nd Quartile Age &  & -0.0920*** \\
             &  & (0.0194) \\
            \addlinespace
            3rd Quartile Age &  & -0.1259*** \\
             &  & (0.0204) \\
            \addlinespace
            4th Quartile Age &  & -0.1194*** \\
             &  & (0.0235) \\
            \addlinespace
            Cloud IaaS (t)$\times$2nd Quartile Age &  & -0.0185 \\
             &  & (0.0450) \\
            \addlinespace
            Cloud IaaS (t)$\times$3rd Quartile Age &  & -0.0459 \\
             &  & (0.0435) \\
            \addlinespace
            Cloud IaaS (t)$\times$4th Quartile Age &  & 0.0178 \\
             &  & (0.0446) \\
            \addlinespace
            Cloud SaaS (t)$\times$2nd Quartile Age &  & -0.0716 \\
             &  & (0.0454) \\
            \addlinespace
            Cloud SaaS (t)$\times$3rd Quartile Age &  & -0.0361 \\
             &  & (0.0429) \\
            \addlinespace
            Cloud SaaS (t)$\times$4th Quartile Age &  & -0.1205*** \\
             &  & (0.0426) \\
            \addlinespace \midrule \addlinespace
            Observations & 16216 & 16216 \\
            Adj R2 & 0.0946 & 0.0902 \\
            Industry, Reg., Year FE & X & X \\
            Controls & X & X \\
            \bottomrule
            \end{tabular}
            \begin{tablenotes}
                \item \textit{Note}: Controls are measured at $t-5$ and include the logarithms of age, tangible and intangible capital, the share of workers in ICT and R\&D occupations, the average hourly wage of managers and engineers, and two dummies for exporter status and multi-establishment firms. Standard errors are clustered at the NACE 2-digit level. The full version of this table, including coefficients for all controls, is available upon request. *** $p<0.01$, ** $p<0.05$, * $p<0.1$.   
            \end{tablenotes}
        \end{minipage}}
\end{threeparttable}
\caption{Robustness -- sales and age quantiles}
\label{tab:robquantiles}
\end{table}

\newpage

\begin{table}[!ht]\centering
\begin{threeparttable}
    \scalebox{0.5}{
        \begin{minipage}{\textwidth}
        \centering
            \begin{tabular}{lcc}
            \toprule
             & (1) & (2) \\
             & Employment \& Age & Employment \& Age, Extended \\
            \addlinespace\cmidrule(lr){2-3}\addlinespace
            Cloud IaaS (t) & 0.1416 & 0.1452 \\
             & (0.1052) & (0.1072) \\
            \addlinespace
            Cloud SaaS (t) & 0.2273* & 0.2198 \\
             & (0.1304) & (0.1332) \\
            \addlinespace
            Size Class = 1 (50-249) & 0.1640*** & 0.1711*** \\
             & (0.0163) & (0.0171) \\
            \addlinespace
            Size Class = 2 (250+) & 0.2902*** &  \\
             & (0.0271) &  \\
            \addlinespace
            Size Class = 3 (250-499) &  & 0.2515*** \\
             &  & (0.0241) \\
            \addlinespace
            Size Class = 4 (500-999) &  & 0.3289*** \\
             &  & (0.0367) \\
            \addlinespace
            Size Class = 5 (1000+) &  & 0.3553*** \\
             &  & (0.0466) \\
            \addlinespace
            Cloud IaaS x Size Class = 1 (50-249) & -0.0021 & -0.0039 \\
             & (0.0302) & (0.0310) \\
            \addlinespace
            Cloud IaaS x Size Class = 2 (250+) & 0.0027 &  \\
             & (0.0251) &  \\
            \addlinespace
            Cloud IaaS x Size Class = 3 (250-499) &  & -0.0325 \\
             &  & (0.0320) \\
            \addlinespace
            Cloud IaaS x Size Class = 4 (500-999) &  & -0.0372 \\
             &  & (0.0330) \\
            \addlinespace
            Cloud IaaS x Size Class = 5 (1000+) &  & 0.0456 \\
             &  & (0.0451) \\
            \addlinespace
            Cloud SaaS x Size Class = 1 (50-249) & -0.0920*** & -0.0848*** \\
             & (0.0287) & (0.0299) \\
            \addlinespace
            Cloud SaaS x Size Class = 2 (250+) & -0.0892*** &  \\
             & (0.0253) &  \\
            \addlinespace
            Cloud SaaS x Size Class = 3 (250-499) &  & 0.0738** \\
             &  & (0.0357) \\
            \addlinespace
            Cloud SaaS x Size Class = 4 (500-999) &  & -0.0730* \\
             &  & (0.0308) \\
            \addlinespace
            Cloud SaaS x Size Class = 5 (1000+) &  & -0.0947* \\
             &  & (0.0372) \\
            \addlinespace
            Age Class = 1 (6-10) & -0.1479*** & -0.1464*** \\
             & (0.0450) & (0.0452) \\
            \addlinespace
            Age Class = 2 (11+) & -0.2687*** &  \\
             & (0.0375) &  \\
            \addlinespace
            Age Class = 3 (11-25) &  & -0.2530*** \\
             &  & (0.0376) \\
            \addlinespace
            Age Class = 4 (25-50) &  & -0.2858*** \\
             &  & (0.0379) \\
            \addlinespace
            Age Class = 5 (50+) &  & -0.2759*** \\
             &  & (0.0390) \\
            \addlinespace
            Cloud IaaS x Age Class = 1 (6-10) & -0.0371 & -0.0353 \\
             & (0.1027) & (0.1039) \\
            \addlinespace
            Cloud IaaS x Age Class = 2 (11+) & -0.1066 &  \\
             & (0.1031) &  \\
            \addlinespace
            Cloud IaaS x Age Class = 3 (11-25) &  & -0.1144 \\
             &  & (0.1079) \\
            \addlinespace
            Cloud IaaS x Age Class = 4 (25-50) &  & -0.1113 \\
             &  & (0.1036) \\
            \addlinespace
            Cloud IaaS x Age Class = 5 (50+) &  & -0.0802 \\
             &  & (0.1067) \\
            \addlinespace
            Cloud SaaS x Age Class = 1 (6-10) & -0.0368 & -0.0350 \\
             & (0.1691) & (0.1713) \\
            \addlinespace
            Cloud SaaS x Age Class = 2 (11+) & -0.1630 &  \\
             & (0.1318) &  \\
            \addlinespace
            Cloud SaaS x Age Class = 3 (11-25) &  & -0.1360 \\
             &  & (0.1885) \\
            \addlinespace
            Cloud SaaS x Age Class = 4 (25-50) &  & -0.1740 \\
             &  & (0.1328) \\
            \addlinespace
            Cloud SaaS x Age Class = 5 (50+) &  & -0.2197 \\
             &  & (0.1326) \\
            \addlinespace \midrule \addlinespace
            Observations & 16216 & 16216 \\
            Adj R2 & 0.112 & 0.116 \\
            Industry, Reg., Year FE & X & X \\
            Controls & X & X \\
            \bottomrule
            \end{tabular}
            \begin{tablenotes}
                \item \textit{Note}: *** $p<0.01$, ** $p<0.05$, * $p<0.1$.
            \end{tablenotes}
        \end{minipage}
    }
\end{threeparttable}
\caption{Robustness -- age and size classes}
\label{tab:robclass}
\end{table}

\newpage
\begin{landscape}

\begin{table}[!htbp] \centering 
		\begin{threeparttable}
			\scalebox{0.65}{
				\begin{minipage}{\textwidth}
					\begin{tabular}{lcccccccccc}
						\toprule
						& (1)        & (2)        & (3)        & (4)        & (5)        & (6)        & (7)        & (8)        & (9)        & (10)       \\
						& t-5;t+1    & t-5;t+2    & t-5;t+3    & t-5;t+4    & t-5;t+5    & t-5;t+1    & t-5;t+2    & t-5;t+3    & t-5;t+4    & t-5;t+5    \\
						\addlinespace\cmidrule(lr){2-11}\addlinespace
						Cloud (t)                             & 0.2687***  & 0.3195***  & 0.3746***  & 0.4124***  & 0.4187***  &            &            &            &            &            \\
						& (0.0871)   & (0.0905)   & (0.0846)   & (0.1035)   & (0.1017)   &            &            &            &            &            \\\addlinespace
						Cloud (t)$\times$Log Sales (t-5)      & -0.0169**  & -0.0219*** & -0.0268*** & -0.0296*** & -0.0305*** &            &            &            &            &            \\
						& (0.0073)   & (0.0075)   & (0.0071)   & (0.0090)   & (0.0087)   &            &            &            &            &            \\\addlinespace
						Cloud IaaS (t)                        &            &            &            &            &            & 0.1363*    & 0.1506*    & 0.2114**   & 0.2388**   & 0.2350**   \\
						&            &            &            &            &            & (0.0728)   & (0.0775)   & (0.0810)   & (0.0926)   & (0.0969)   \\\addlinespace
						Cloud SaaS (t)                        &            &            &            &            &            & 0.2507**   & 0.3109***  & 0.3025***  & 0.3034***  & 0.3216***  \\
						&            &            &            &            &            & (0.0964)   & (0.1048)   & (0.0826)   & (0.1088)   & (0.0906)   \\\addlinespace
						Cloud IaaS (t)$\times$Log Sales (t-5) &            &            &            &            &            & -0.0061    & -0.0084    & -0.0135*   & -0.0156*   & -0.0145    \\
						&            &            &            &            &            & (0.0069)   & (0.0071)   & (0.0075)   & (0.0084)   & (0.0089)   \\\addlinespace
						Cloud SaaS (t)$\times$Log Sales (t-5) &            &            &            &            &            & -0.0208**  & -0.0257*** & -0.0251*** & -0.0252*** & -0.0285*** \\
						&            &            &            &            &            & (0.0081)   & (0.0090)   & (0.0070)   & (0.0090)   & (0.0077)   \\\addlinespace
						Log Sales (t-5)                       & -0.0342*** & -0.0234**  & -0.0280*** & -0.0246**  & -0.0305*** & -0.0327*** & -0.0216**  & -0.0265*** & -0.0231**  & -0.0286*** \\
						& (0.0067)   & (0.0091)   & (0.0097)   & (0.0102)   & (0.0096)   & (0.0066)   & (0.0089)   & (0.0096)   & (0.0101)   & (0.0095)   \\
						\addlinespace\midrule\addlinespace
						Observations                          & 15,802     & 15,348     & 14,734     & 14,491     & 13,952     & 15,802     & 15,348     & 14,734     & 14,491     & 13,952     \\
						Adj R2                                & 0.0949     & 0.117      & 0.113      & 0.131      & 0.138      & 0.0959     & 0.118      & 0.114      & 0.132      & 0.139      \\
						Industry, Reg., Year FE               & X          & X          & X          & X          & X          & X          & X          & X          & X          & X          \\
						Firm controls                         & X          & X          & X          & X          & X          & X          & X          & X          & X          & X         \\\bottomrule
					\end{tabular}
					\begin{tablenotes}
						\item \textit{Note}: The table reports the results of the estimation of Equation \ref{eq:baseline_OLS}. Controls are measured at $t-5$ and include the logarithms of age, tangible and intangible capital, the share of workers in ICT and R\&D occupations, the average hourly wage of managers and engineers, and two dummies for exporter status and multi-establishment firms. Standard errors are clustered at the NACE 2-digit level. The full version of this table, including coefficients for all controls, is available upon request. *** $p<0.01$, ** $p<0.05$, * $p<0.1$.   
					\end{tablenotes}
				\end{minipage}
			}
		\end{threeparttable}
            \caption{Forward growth rates}
            \label{tab:robforwardgrowth}
	\end{table}
\end{landscape}

\newpage

\begin{table}[!htbp]\centering
    \begin{threeparttable}
        \scalebox{0.65}{\color{black}
            \begin{minipage}{\textwidth}
                \centering
                \begin{tabular}{lcccccc}
                    \toprule
                    & (1)        & (2)        & (3)        & (4)        & (5)        & (6)        \\
                    & \multicolumn{2}{c}{2016-2019} & \multicolumn{2}{c}{2018-2019} & \multicolumn{2}{c}{2017-2019} \\\addlinespace\cmidrule(lr){2-3}\cmidrule(lr){4-5}\cmidrule(lr){6-7}\addlinespace
                    Cloud (t)                            & 0.0985**   &            & 0.0195     &            & 0.0570*    &            \\
                    & (0.0487)   &            & (0.0162)   &            & (0.0286)   &            \\\addlinespace
                    IaaS Cloud (t)                       &            & 0.0293     &            & -0.0080    &            & 0.0270     \\
                    &            & (0.0598)   &            & (0.0220)   &            & (0.0346)   \\\addlinespace
                    SaaS Cloud (t)                       &            & 0.1205*    &            & 0.0281     &            & 0.0485*    \\
                    &            & (0.0716)   &            & (0.0283)   &            & (0.0277)   \\\addlinespace
                    Lag Sales                            & 0.0114**   & 0.0117**   & -0.0000    & -0.0001    & 0.0038     & 0.0041     \\
                    & (0.0047)   & (0.0047)   & (0.0019)   & (0.0019)   & (0.0030)   & (0.0029)   \\\addlinespace
                    Cloud (t)$\times$Lag Sales           & -0.0086**  &            & -0.0015    &            & -0.0046*   &            \\
                    & (0.0042)   &            & (0.0016)   &            & (0.0027)   &            \\\addlinespace
                    IaaS Cloud (t)$\times$Lag Sales      &            & -0.0014    &            & 0.0008     &            & -0.0016    \\
                    &            & (0.0054)   &            & (0.0021)   &            & (0.0032)   \\\addlinespace
                    SaaS Cloud (t)$\times$Lag Sales      &            & -0.0120*   &            & -0.0023    &            & -0.0047*   \\
                    &            & (0.0063)   &            & (0.0025)   &            & (0.0026)   \\
                    \addlinespace\midrule\addlinespace
                    Observations                          & 8,333      & 8,333      & 7,841      & 7,841      & 7,855      & 7,855      \\
                    Adj R2                                 & 0.0332     & 0.0336     & 0.0253     & 0.0250     & 0.0309     & 0.0309     \\
                    Industry, Reg., Year FE                           & X          & X          & X          & X          & X          & X          \\
                    Other Controls                                & X          & X          & X          & X          & X          & X          \\\bottomrule
                \end{tabular}
                \begin{tablenotes}
                    \footnotesize
						\item \textit{Note}: The table reports the results of the estimation of Equation \ref{eq:baseline_OLS}. Growth rates are computed between 2016 and 2019 in Column 1 and 2, in 2018-2019 in Column 3 and in 2017-2019 in Column 4. Models are estimated on the 2016 ICT survey in Columns 1 and 2 and on the 2018 ICT survey in Columns 3 and 4. Controls are measured at $t-5$ and include the logarithms of age, tangible and intangible capital, the share of workers in ICT and R\&D occupations, the average hourly wage of managers and engineers, and two dummies for exporter status and multi-establishment firms. Standard errors are clustered at the NACE 2-digit level. The full version of this table, including coefficients for all controls, is available upon request. *** $p<0.01$, ** $p<0.05$, * $p<0.1$.   
                \end{tablenotes}
            \end{minipage}}
    \end{threeparttable}
    \caption{Cloud adoption and forward firm sales growth}
    \label{tab:forward2}
\end{table}

\newpage

\begin{table}[!htbp] \centering
    \begin{threeparttable}
        \scalebox{0.65}{
            \begin{minipage}{\textwidth}
            \begin{tabular}{lcc}
						\toprule
						& (1)          & (2)          \\
						& Growth t;t-3 & Growth t;t-7 \\
						\addlinespace\cmidrule(lr){2-3}\addlinespace
						Cloud IaaS (t)                        & 0.0044       & 0.1753*      \\
						& (0.0664)     & (0.1007)     \\\addlinespace
						Cloud SaaS (t)                        & 0.1892**     & 0.3179**     \\
						& (0.0728)     & (0.1225)     \\\addlinespace
						Cloud IaaS (t)$\times$Log Sales (t-5) & 0.0019       & -0.0080      \\
						& (0.0058)     & (0.0092)     \\\addlinespace
						Cloud SaaS (t)$\times$Log Sales (t-5) & -0.0158**    & -0.0255**    \\
						& (0.0065)     & (0.0105)     \\\addlinespace
						Log Sales (t-5)                       & -0.0182**    & -0.0514***   \\
						& (0.0070)     & (0.0078)     \\
						\addlinespace\midrule\addlinespace
						Observations                          & 17,327       & 15,270       \\
						Industry, Reg., Year FE               & X            & X            \\
						Controls                              & X            & X            \\
						Adj. R2                               & 0.0584       & 0.117      \\
						\bottomrule
					\end{tabular}
                \begin{tablenotes}
                    \small
                    \item \textit{Note}: Change in length of growth rate. Column 1 is t;t-3, Column 2 is Growth t;t-7. Controls are measured at $t-5$ and include the logarithms of age, tangible capital, the share of workers in ICT and R\&D occupations, the average hourly wage of managers and engineers, and two dummies for exporter status and multi-establishment firms. Controls in Column (1) also include intangible capital. Standard errors are clustered at the NACE 2-digit level. The full version of this table, including coefficients for all controls, is available upon request. *** $p<0.01$, ** $p<0.05$, * $p<0.1$.   
                \end{tablenotes}
            \end{minipage}
        }
    \end{threeparttable}
    \caption{Change in growth rate length}
    \label{tab:roblength}
\end{table}

\begin{table}[!htbp] \centering
    \begin{threeparttable}
        \scalebox{0.65}{
            \begin{minipage}{\textwidth}
            \begin{tabular}{lcc}
            \toprule
                                      & (1)        & (2)        \\
                                      \addlinespace\cmidrule(lr){2-3}\addlinespace
Cloud IaaS (t)                        & 0.1290     & 0.2160**   \\
                                      & (0.1553)   & (0.1041)   \\\addlinespace
Cloud SaaS (t)                        & 0.4507***  & 0.3014**   \\
                                      & (0.1591)   & (0.1324)   \\\addlinespace
Cloud IaaS (t)$\times$Log Sales (t-5) & -0.0021    & -0.0125    \\
                                      & (0.0148)   & (0.0091)   \\\addlinespace
Cloud SaaS (t)$\times$Log Sales (t-5) & -0.0371**  & -0.0248**  \\
                                      & (0.0142)   & (0.0112)   \\\addlinespace
Log Sales (t-5)                       & -0.0507*** & -0.0419*** \\
                                      & (0.0118)   & (0.0090)   \\
                                      \addlinespace\midrule\addlinespace
Observations                          & 7,207      & 8,019      \\
Adj R2                                & 0.124      & 0.119      \\
Industry, Reg., Year FE               & X          & X          \\
Controls                              & X          & X         \\
\bottomrule
                \end{tabular}
                \begin{tablenotes}
                    \small
                    \item \textit{Note}: The table reports the results of the estimation of Equation \ref{eq:longdiff}. Column 1 and 2 report the results of the Long difference regressions estimated with the 2018 and 2016 ICT surveys. Controls are measured at $t-5$ and include the logarithms of age, tangible capital, the share of workers in ICT and R\&D occupations, the average hourly wage of managers and engineers, and two dummies for exporter status and multi-establishment firms. Standard errors are clustered at the NACE 2-digit level. The full version of this table, including coefficients for all controls, is available upon request. *** $p<0.01$, ** $p<0.05$, * $p<0.1$.   
                \end{tablenotes}
            \end{minipage}
        }
    \end{threeparttable}
    \caption{Long difference regressions}
    \label{tab:roblongdiff}
\end{table}

\end{document}